\documentclass[preprint]{elsarticle}
\usepackage{epsfig}
\usepackage[english]{babel}
\usepackage{amssymb}

\begin{document}
\begin{frontmatter}
\title{The beam energy measurement system for the Beijing
electron-positron collider.}  
\author[binp]{E.V.~Abakumova}
\author[binp]{M.N.~Achasov\corref{cor}}
\ead{achasov@inp.nsk.su}
\author[binp]{V.E.~Blinov}
\author[ihep]{X.~Cai}
\author[ihep]{H.Y.~Dong}
\author[ihep]{C.D.~Fu}
\author[huni]{F.A.~Harris}
\author[binp]{V.V.~Kaminsky}
\author[binp]{A.A.~Krasnov}
\author[huni]{Q.~Liu}
\author[ihep]{X.H.~Mo}
\author[binp]{N.Yu.~Muchnoi}
\author[binp]{I.B.~Nikolaev}
\author[ihep]{Q.~Qin}
\author[ihep]{H.M.~Qu}
\author[huni]{S.L.~Olsen}
\author[binp]{E.E.~Pyata}
\author[binp]{A.G.~Shamov}
\author[huni]{C.P.~Shen}
\author[binp]{K.Yu.~Todyshev}
\author[huni]{G.S.~Varner}
\author[ihep]{Y.F.~Wang}
\author[ihep]{Q.~Xiao}
\author[ihep]{J.Q.~Xu}
\author[ihep]{J.Y.~Zhang}
\author[ihep]{T.B.~Zhang}
\author[ihep]{Y.H.~Zhang}
\author[binp]{A.A.~Zhukov}

\cortext[cor]{Corresponding author}
\address[binp]{Budker Institute of Nuclear Physics, Siberian Branch of the
Russian Academy of Sciences, 11 Lavrentyev,
Novosibirsk 630090, Russia}
\address[ihep]{Institute of High Energy Physics, Beijing 100049, 
People's Republic of China}
\address[huni]{University of Hawaii, Honolulu, Hawaii, 96822, USA}

\begin{abstract}
 The beam energy measurement system (BEMS) for the upgraded Beijing
 electron-positron collider BEPC-II is described. The system is based
 on measuring the energies of Compton back-scattered photons. The
 relative systematic uncertainty of the electron and positron beam
 energy determination is estimated as $2\cdot10^{-5}$. The relative
 uncertainty of the beam's energy spread is about 6\%.
\end{abstract}

\end{frontmatter}

\section{Introduction}

 The upgraded Beijing electron-positron collider (BEPC-II) is a
 $\tau$-charm factory with a center of mass energy range from 2.0 to
 4.6 GeV and a design peak luminosity of $10^{33}$ cm$^{-2}$ s$^{-1}$
 \cite{bepc2}.  For experiments at BEPC-II, the BESIII (Beijing
 spectrometer) detector with high efficiency and resolution both for
 charged and neutral particles was constructed \cite{bes3}. BESIII
 started data taking in 2008. The BESIII research program covers
 charmonium physics, $D$-meson physics, spectroscopy of light hadrons
 and $\tau$-lepton physics \cite{yellbook}.  The $\tau$-lepton is a
 fundamental particle, and its mass is a Standard Model parameter,
 which requires that its mass be determined with high precision. The
 measurements of the $\psi$ and $D$ meson masses are also of
 interest.

 The current value of the $\tau$ mass, $m_\tau$, is $1776.82\pm 0.16$
 \cite{PDG}. In BEPC-II/BESIII, the mass will be measured using the
 threshold scan method. The accuracy of the measurement was studied in
 Ref.~\cite{mo1,mo2}. Two weeks of data taking will lead to a
 statistical uncertainty of less then 50 keV. The systematic
 uncertainty (without the accuracy of beam energy determination) is
 about 20 keV and includes uncertainties of the luminosity, detection
 efficiency, branching fraction, background, energy spread, and
 theoretical uncertainty. The most important source of uncertainty is
 the accuracy of the absolute beam energy determination.
 
 In some cases, the energy scale of colliders can be calibrated with
 extremely high accuracy using the resonant depolarization technique
 \cite{rd}.  But this approach is not applicable for $e^+e^-$
 factories, where the great advance in luminosity is made possible by
 fast bunch-to-bunch feedback systems that usually have a strong
 depolarization impact on the beam. There are two possible methods of
 the beam energy determination at BEPC-II. First is a calibration of
 the energy scale from scan of the $J/\psi$ and $\psi^\prime$
 resonances \cite{bes-tau}. The expected accuracy in this case is
 about 100 keV.
 
 Another possibility is the beam energy measurement using Compton
 back-scattering of monochromatic laser radiation on the $e^\pm$ beams.
 This approach was developed and experimentally proved in
 Ref.~\cite{okp1,okp2,okp3,okp4}. At the BESSY-I and BESSY-II storage
 rings, the relative accuracies of energy measurement of about
 $10^{-4}$ and $3\times 10^{-5}$ for the beam energies of 800 and 1700
 MeV, respectively, were achieved \cite{okp3}. In collider experiments,
 this method was applied at VEPP-4M \cite{okp4}. Based on the VEPP-4M
 experience, such a system was proposed and constructed for BEPC-II
 \cite{proposal,prop}.  In this paper, the system design and
 performance are reported.

\section{The Compton Back-scattering approach}

\begin{figure}
\begin{center}
\includegraphics[scale=0.5]{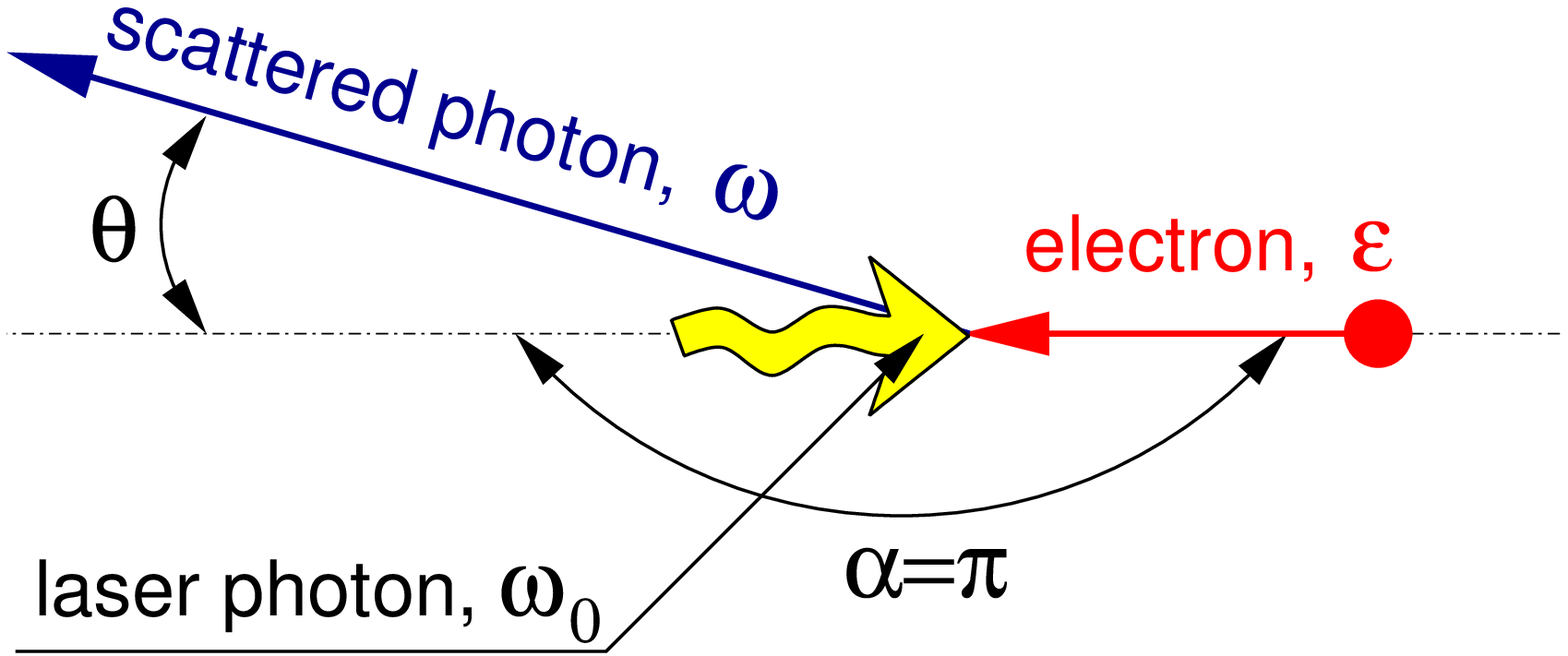}
\caption{The Compton scattering process. $\varepsilon$, $\omega_0$, and
         $\omega$ are the particles energies, and $\alpha=\pi$.}
\label{cs-kin}
%\end{figure}
%\begin{figure}
%\centering
\includegraphics[scale=0.5]{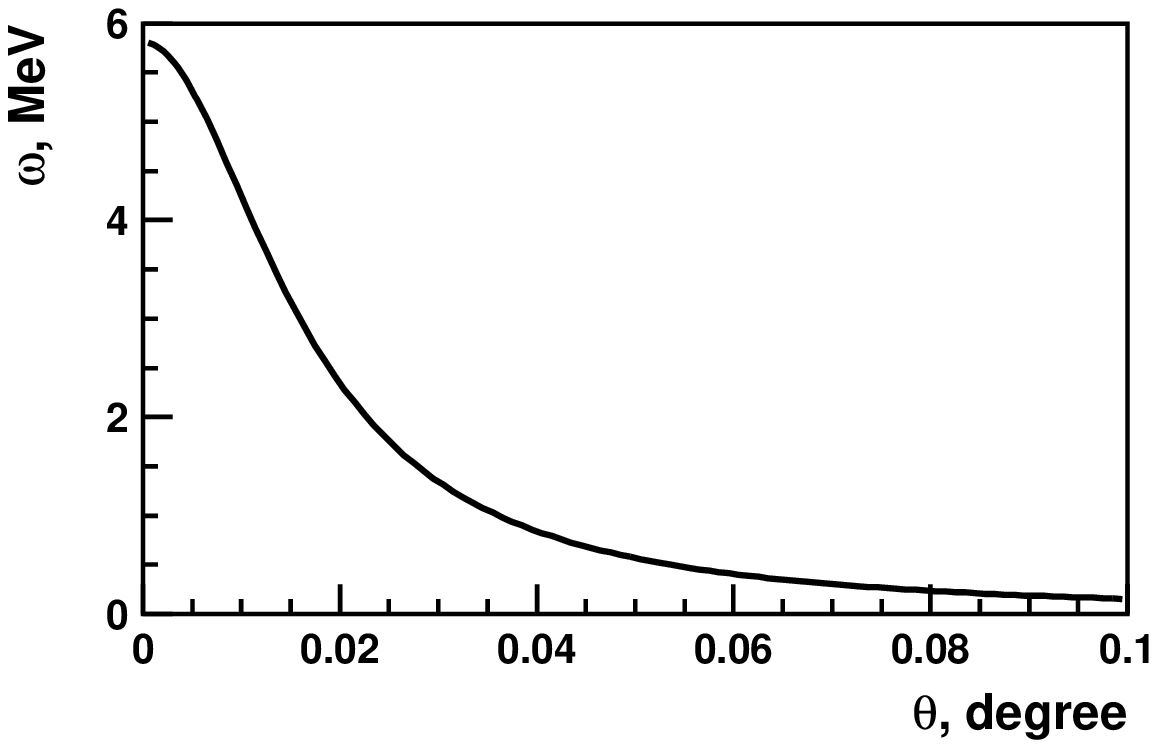}
\caption{The dependence of the scattered photon energy $\omega$ on the
         angle $\theta$ between the initial electron and the final
         photon in the Compton scattering process. The initial
         electron and photon energies are $\omega_0=0.12$ eV and
         $\varepsilon=1770$ MeV, respectively, and $\alpha=\pi$.}
\label{cs-y-e}
%\end{figure}
%\begin{figure}
\centering
\includegraphics[scale=0.5]{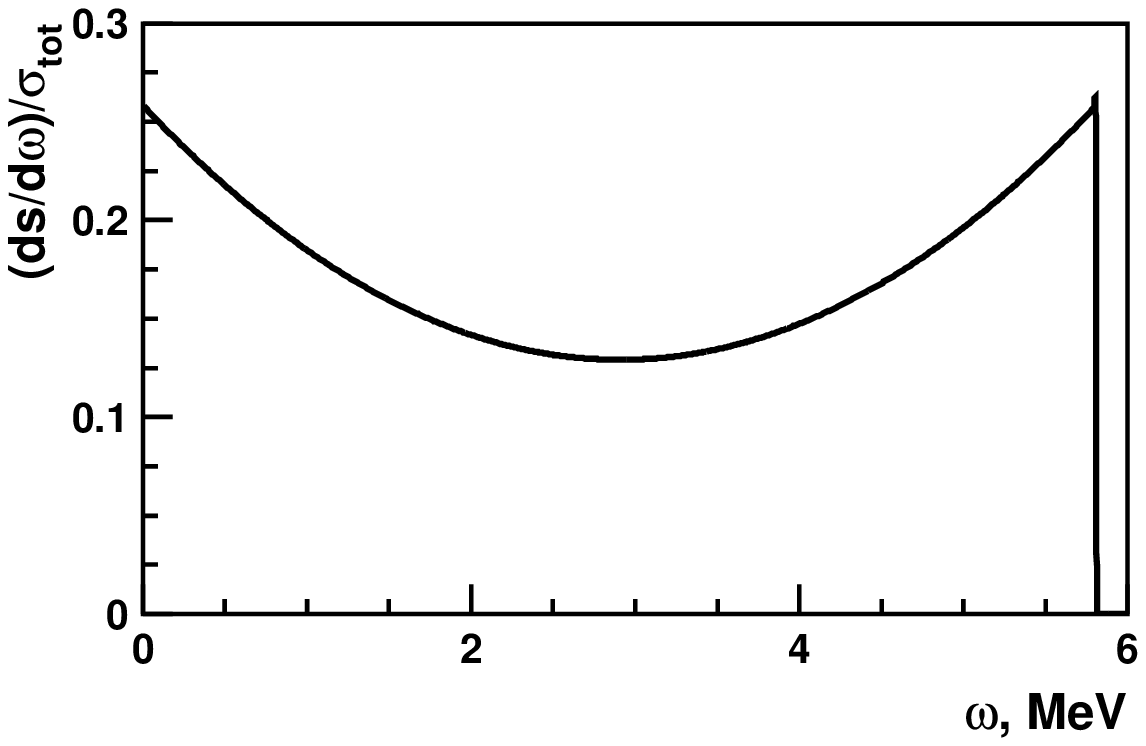}
\caption{Energy spectrum of scattered Compton photons. The initial electron
         and photon energies are $\omega_0=0.12$ eV and $\varepsilon=1770$ 
	 MeV, respectively, and $\alpha=\pi$.}
\label{cs-e-cnek}
\end{center}
\end{figure}

 Let us consider the Compton scattering process in a case where the
 angle $\alpha$ between initial particles is equal to $\pi$ and their
 energies are $\omega_0\ll m_e\ll \varepsilon$
 (Fig.~\ref{cs-kin}). Here $\omega_0$ and $\varepsilon$ are the
 energies of the initial photon and electron, respectively.  The
 back-scattered photons with $\theta=0$ have the maximal energy
 (Fig.~\ref{cs-y-e}), and the energy spectrum of the scattered photons
 has a sharp edge at the maximal energy (Fig.~\ref{cs-e-cnek}).

The general idea is based on the following:
\begin{itemize}
\item 
 The maximal energy of the scattered photon $\omega_{max}$ is related with the
 electron energy $\varepsilon$ by the kinematics of Compton scattering
 \cite{proposal}:
\begin{equation}
 \omega_{max}={{\varepsilon^2}\over{\varepsilon+m_e^2/4\omega_0}},
\end{equation}
 If one measures $\omega_{max}$, then the electron energy can be calculated:
\begin{equation} 
 \varepsilon={\omega_{max}\over 2} \Biggl[ 1 +
 \sqrt{1+{ m_e^2 \over {\omega_0\omega_{max}} } }\;\Biggr].\label{nepec}
\end{equation}
\item
 The ultra-high energy resolution ($\sim 10^{-3}$) of commercially 
 available High Purity Germanium (HPGe) detectors allows the statistical 
 accuracy in the  beam energy measurement to be at the level of 
 $\delta\varepsilon/\varepsilon \simeq 10^{-5}$.
\item
 The systematical accuracy is mostly defined by absolute calibration of the
 detector energy scale. Accurate calibration can be performed  in the
 photon energy range up to about 10~MeV by using $\gamma$-active radionuclides.
\end{itemize}
\begin{figure}[p]
\centering
\includegraphics[scale=0.7]{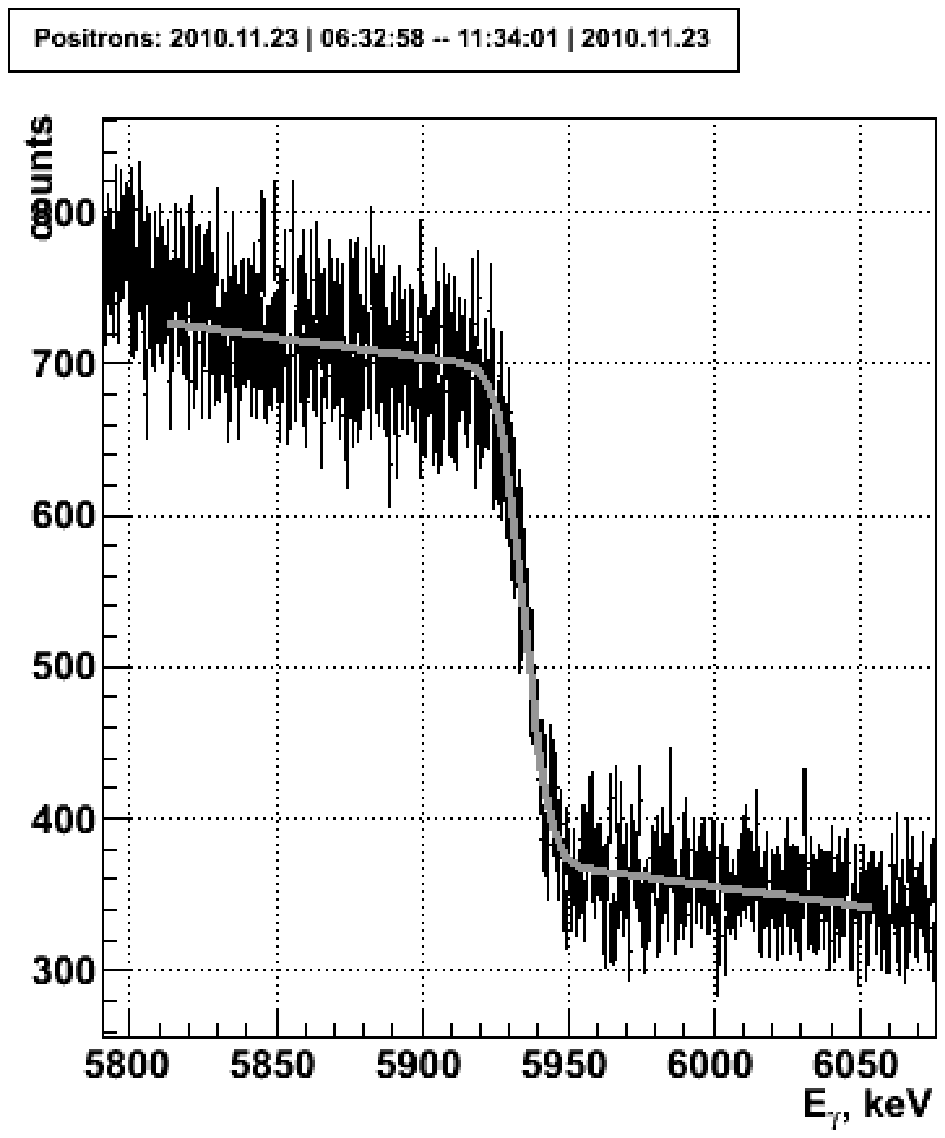}
\caption{The measured edge of the scattered photons energy spectrum. The line
         is the fit result.}
\label{edge}
\includegraphics[scale=0.7]{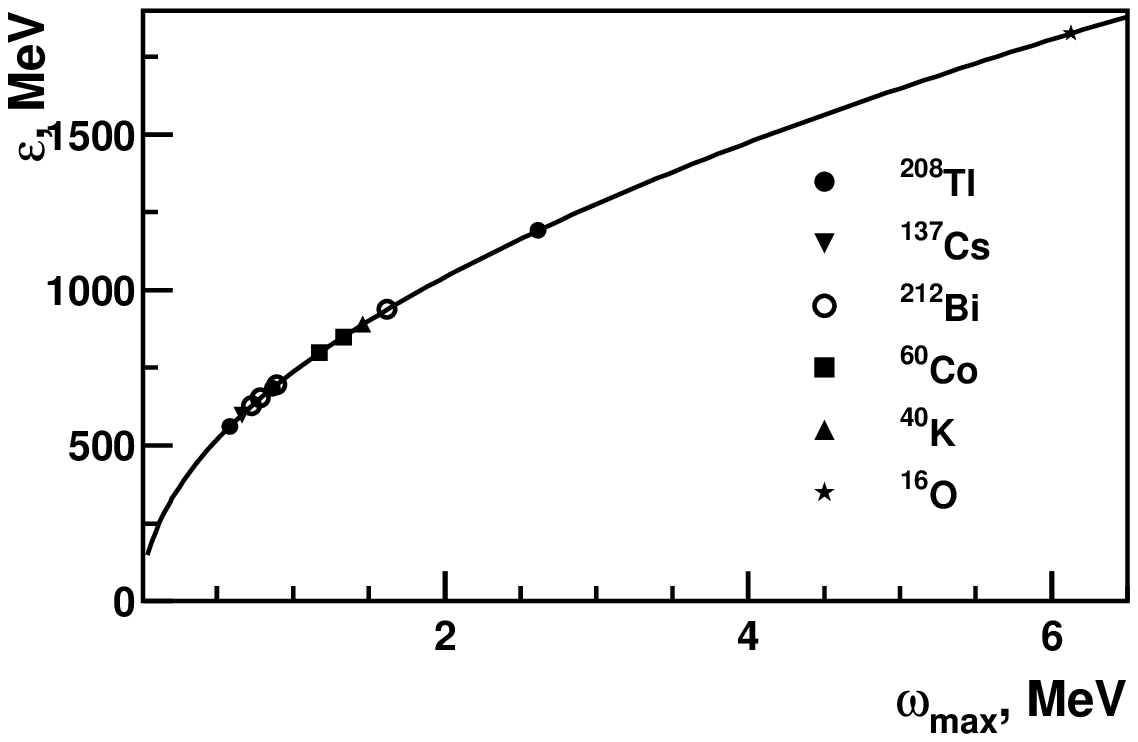}
\caption{Relation between $\omega_{max}$ and $\varepsilon$ (solid line). Dots
         are the energies of $\gamma$-active radionuclide reference lines
	 for the HPGe detector calibration. The initial photon energy is 
	 $\omega_0=0.12$ eV.}
\label{lines}
\end{figure}

 The measurement procedure is as follows. As a source of initial
 photons, the monochromatic laser radiation with $\omega_0\approx
 0.12$~eV is used. The laser light is put in collision with the
 electron or positron beams, and the energy of the back-scattered
 photons is precisely measured using the HPGe detector.  The maximal
 energy of the scattered photons is determined by fitting the abrupt
 edge in the energy spectrum by the erfc-like function
 (Fig.~\ref{edge}). The relation between the measured $\omega_{max}$
 and the beam energy $\varepsilon$ is shown in Fig.~\ref{lines}. The
 detector energy scale is accurately calibrated by using well-known
 radiative sources of $\gamma$-radiation (Fig.~\ref{lines}).

\section{The beam energy measurement system for BEPC-II.}
  
\begin{figure}
\begin{center}
\includegraphics[scale=0.5]{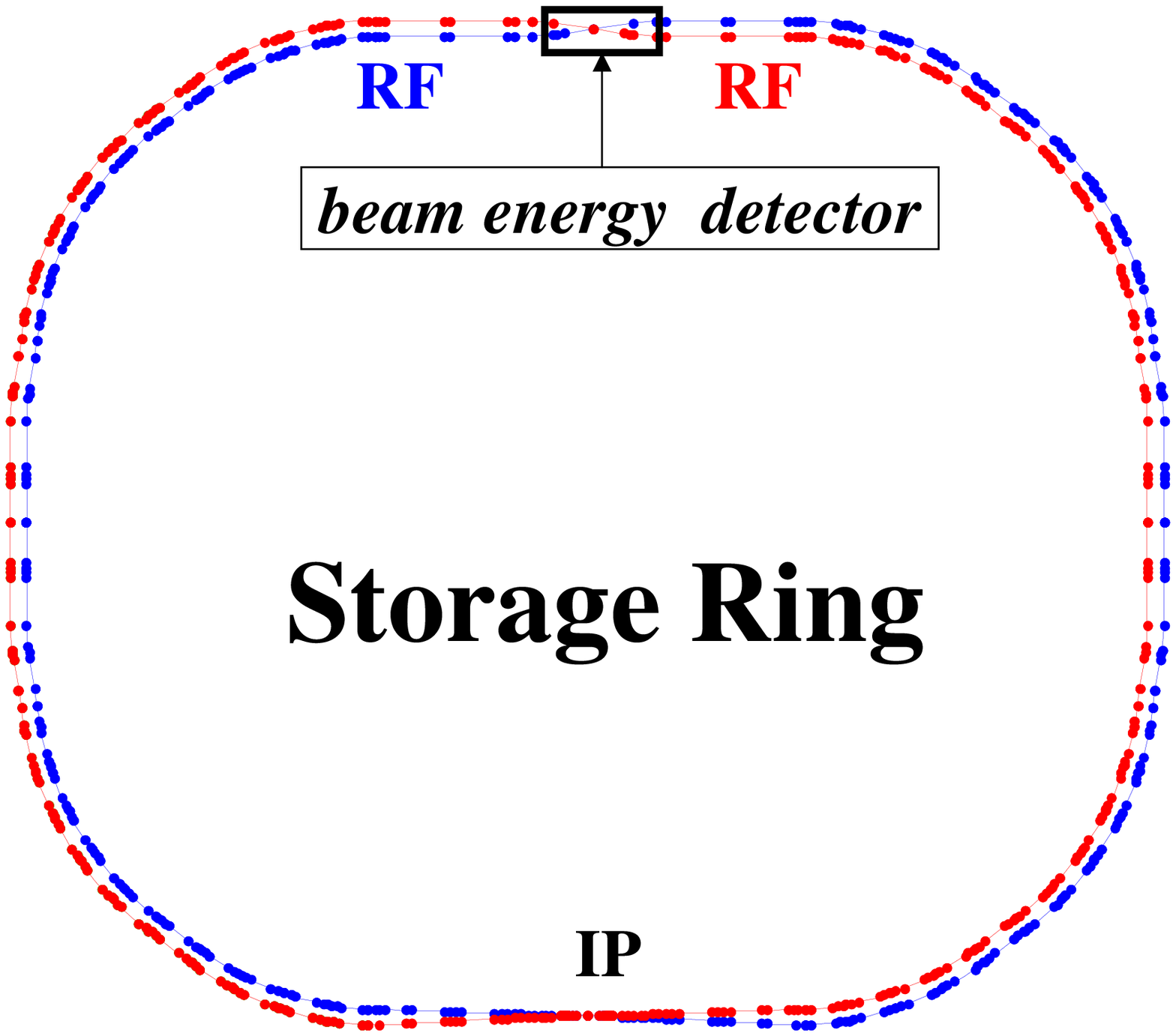}
\caption{Location of the energy measurement system at the BEPC-II collider.
         The deployment place is indicated as ``beam energy detector''.}
\label{bepcsr}
%\end{figure*}
%\begin{figure*}
%\centering
\includegraphics[scale=0.5]{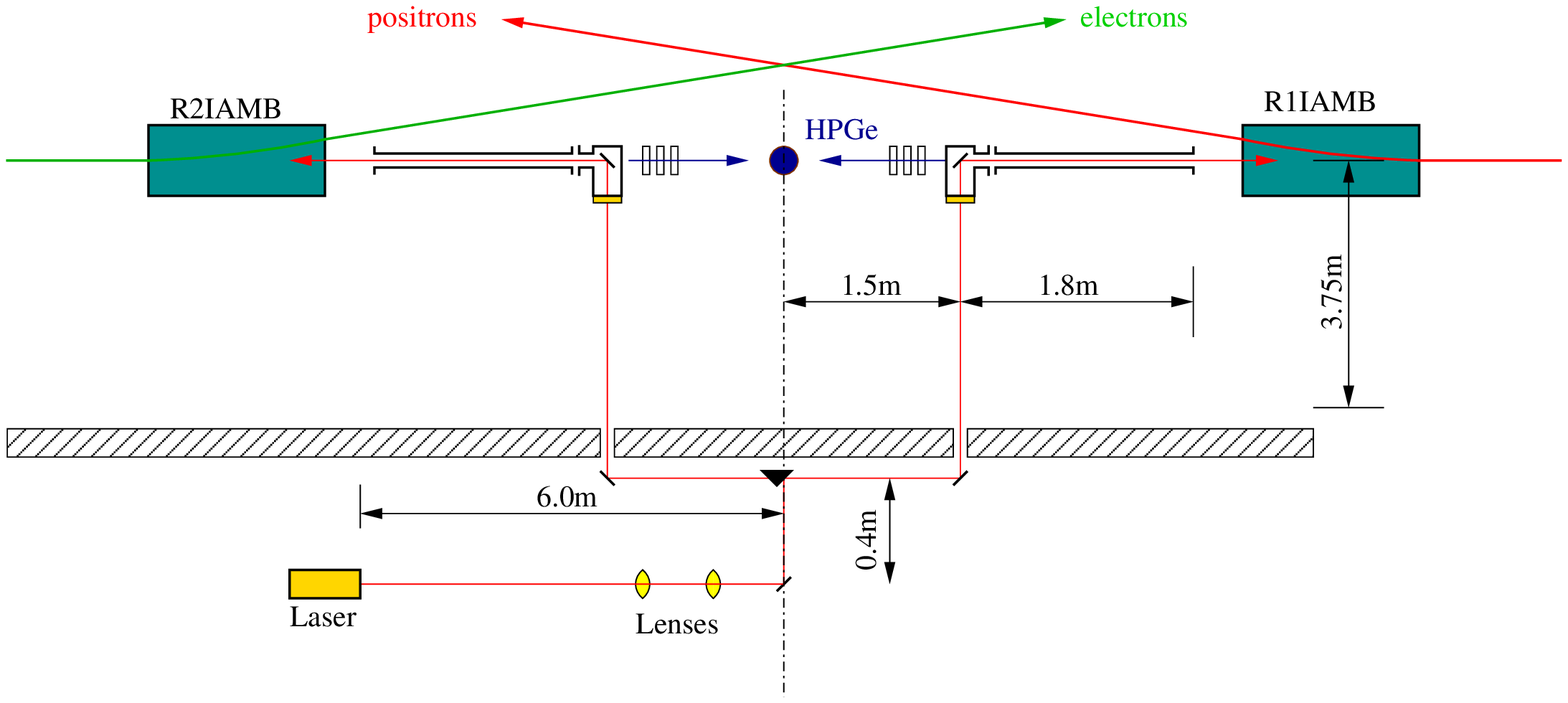}
\caption{Simplified schematic of the energy measurement system. The positron
         and electron beams are indicated. R1IAMB and R2IAMB are accelerator 
	 magnets, and the HPGe detector is represented by the dot at the 
	 center. The shielding wall of the beam tunnel is shown cross-hatched,
	 and the laser is located outside the tunnel.}
\label{layout}
\end{center}
\end{figure}

 The beam energy measurement system is located at the north beam crossing 
 point of the BEPC-II storage rings (Fig.~\ref{bepcsr}). This location allows
 measurement of the electron and positron beams energy by the same HPGe
 detector. The layout schematic of the system is shown in Fig.~\ref{layout}.

 The system consists of the laser source, optical and laser-to-vacuum
 insertion systems to transport the laser beam into the interaction
 regions where the laser beam collides with either the electron or
 positron beam, and the HPGe detector to measure back-scattered
 photons. The laser and optical system elements are deployed in the
 corridor outside the collider hall.
 
 The laser and electron (positron) beams interact in the straight sections of
 the collider's rings beyond the R2IAMB (R1IAMB) dipole magnets. The total 
 yield of scattered photons was estimated in Ref.~\cite{proposal} and is 
 about 17000 gammas per second, per 1 mA of electron (positron) beam 
 current, per 1 W of laser power.

\subsection{Laser and optical system}
 
 The source of initial photons is the GEM Selected
 50$^{\mbox{\scriptsize TM}}$ $CO_2$ laser from Coherent, {\em
 Inc.}. It is a continuous operation (CW), high power, and single-line
 laser. It provides 25 W of CW power at the wavelength
 $\lambda_0=10.835231$ $\mu$m ( $\gamma$-quantum energy
 $\omega_0=0.114426901$ eV), which corresponds to the 10P42 transition
 in the carbon dioxide molecule \cite{CO2lambda}. $\omega_0$ is
 known with relative accuracy better then 0.1~ppm. The relative width
 of the laser photon spectrum is $\sigma_\omega/\omega_0\approx$
 3~ppm.  This wavelength was adopted in order to avoid any
 interference between $\gamma-$radiation lines of radiative sources,
 used for HPGe detector calibration, and the Compton edges of all
 interesting energy points in the $\tau$-charm energy region
 (Fig.~\ref{laser-kpa}).  The laser is installed on a special support
 which can be adjusted as necessary.
 
 The optical system includes the following units which are situated along the
 collider wall (Fig.~\ref{layout}):
\begin{enumerate}
\item
 Two ZnSe lenses with focal lengths of $f=40$ cm. The laser beam is
 focused at the BEPC-II vacuum chamber entrance
 flange, where the geometrical aperture is minimal: vertical size
 $\times$ horizontal size is 14 mm $\times$ 50 mm. The total distance
 from the laser output aperture to the entrance flange of the BEPC-II
 vacuum chamber is about 18 m. The lenses are placed at 300 and 382 cm
 from the laser and provide the laser beam transverse size at the
 flange from 0.20 to 0.25 cm.
\item 
A $45^\circ$ mirror, which reflects the beam through an angle of
 $90^\circ$ towards the movable prism.
\item
 A movable reflector prism which directs the laser beam towards the
 right or left mirror.
\item
 Two mirrors to reflect the right or left-traveling laser beam into
 the collider tunnel through holes in the concrete wall. The laser
 beam is incident on a viewport in a vacuum pipe extension of the beam
 pipe.  The mirrors are installed on special supports that allow
 precise vertical and horizontal angular alignment by the use of
 stepping motors (one step equals $1.5\times 10^{-6}$ rad).
\end{enumerate}

\subsection{Laser-to-vacuum insertion system}

\begin{figure}
\begin{center}
\includegraphics[scale=0.5]{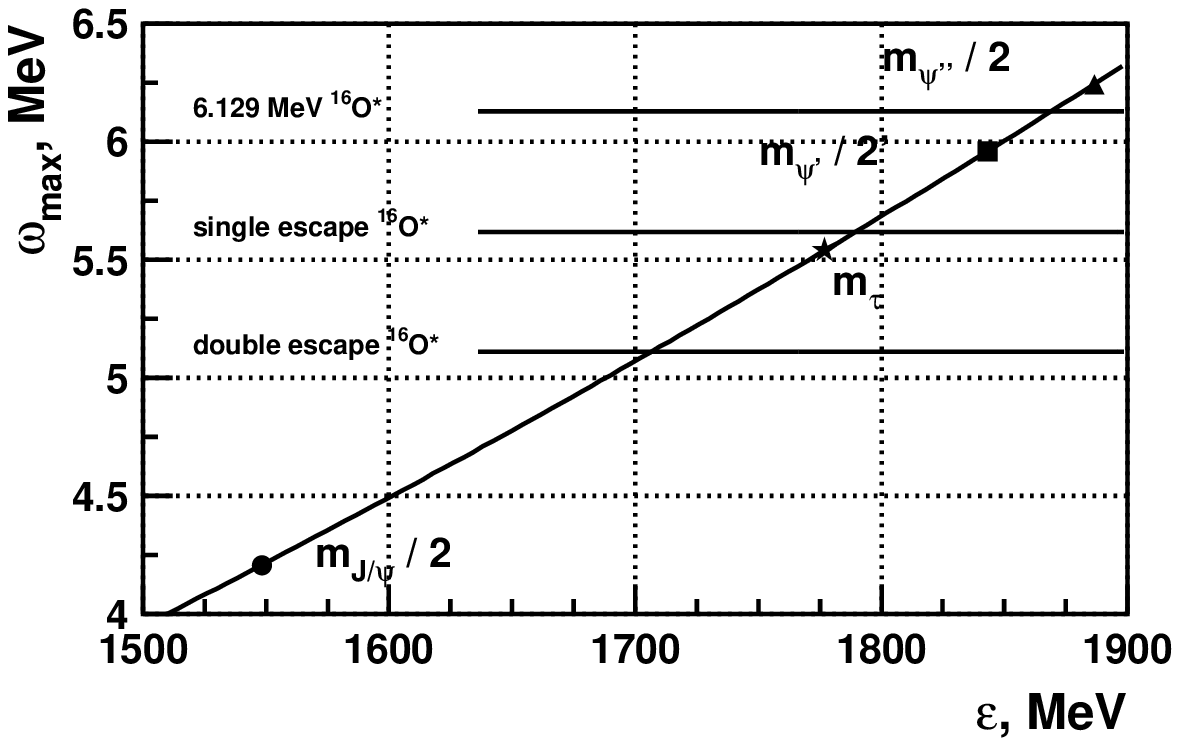}
\caption{The relation between beam energy $\varepsilon$ and energy of 
         back-scattered Compton photons $\omega_{max}$. The $\gamma$ lines of 
	 $^{16}O^*$ are also shown. }
\label{laser-kpa}
%\end{figure*}
%\begin{figure*}
\includegraphics[scale=0.3]{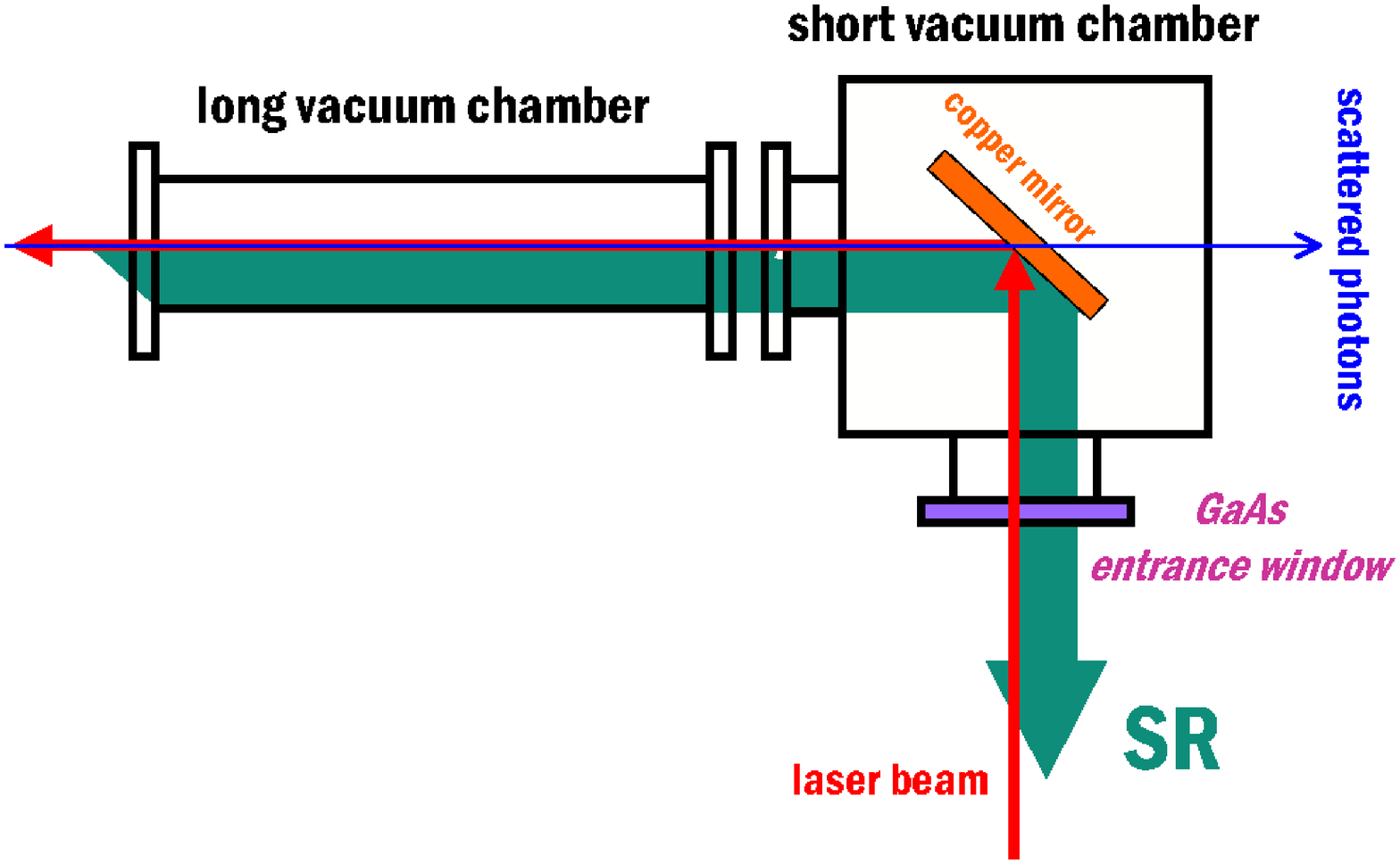}
%\vspace{-1cm}
\caption{Simplified schematic of the laser-to-vacuum insertion assembly.}
\label{cxe-bakyy}
\includegraphics[scale=0.15]{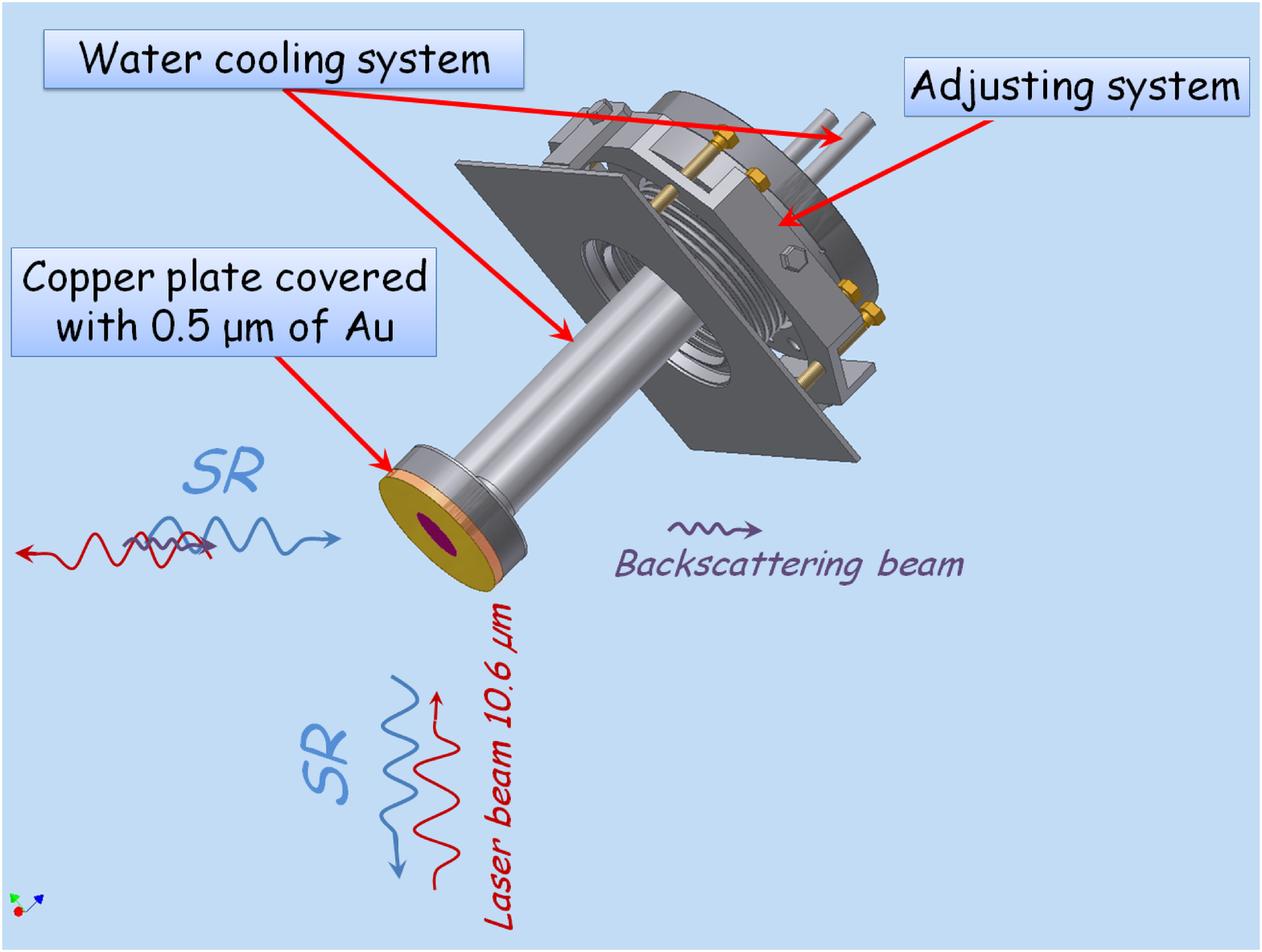}
\caption{Copper mirror.}
\label{mirror}
\end{center}
\end{figure}

 The insertion of the laser beam into the vacuum chamber is performed
 using the laser-to-vacuum insertion system. The system is the special
 stainless steel vacuum chamber with a GaAs entrance viewport
 \cite{win1} and water cooled copper mirror (Fig.~\ref{cxe-bakyy}). In
 the vacuum chamber, the laser beam is reflected through an angle of
 $90^\circ$ by the copper mirror.  After back-scattering, the photons
 return to the mirror, pass through it, leave the vacuum chamber, and
 are detected by the HPGe detector. Note, the copper mirror protects
 the view port against high power synchrotron radiation due to low
 reflectivity of high energy photons (less than 1\%) from a metallic
 surface.

 The copper mirror design is shown in Fig.~{\ref{mirror}}. The mirror
 can be turned by bending the vacuum flexible bellows, so the angle
 between the mirror and the laser can be adjusted as
 necessary. Synchrotron radiation (SR) photons heat the mirror.  In
 order to reduce the heating of the mirror, it is placed 1.8 m from
 the BEPC-II vacuum chamber flange. The SR power absorbed by the
 mirror is about 200 W. The extraction of heat is provided by a
 water cooling system. To prevent adsorption of residual gas
 molecules on the mirror surface, it is covered with a 0.5 $\mu$m
 thick gold layer.

\begin{figure}
\begin{center}
\includegraphics[scale=0.2]{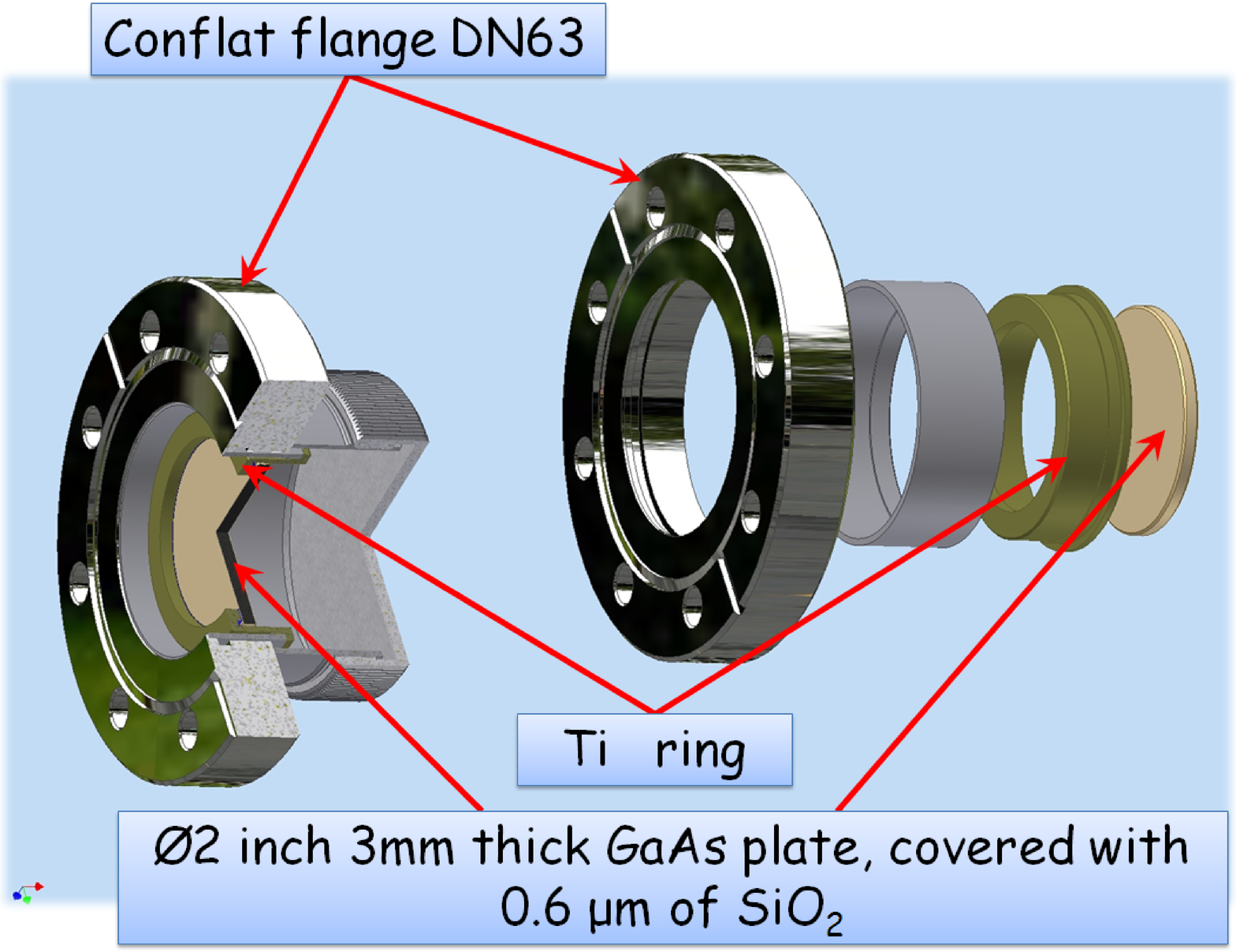}
\caption{The $GaAs$ viewport.}
\label{viewport}
%\end{figure*}
%\begin{figure*}
%\centering
\includegraphics[scale=0.4]{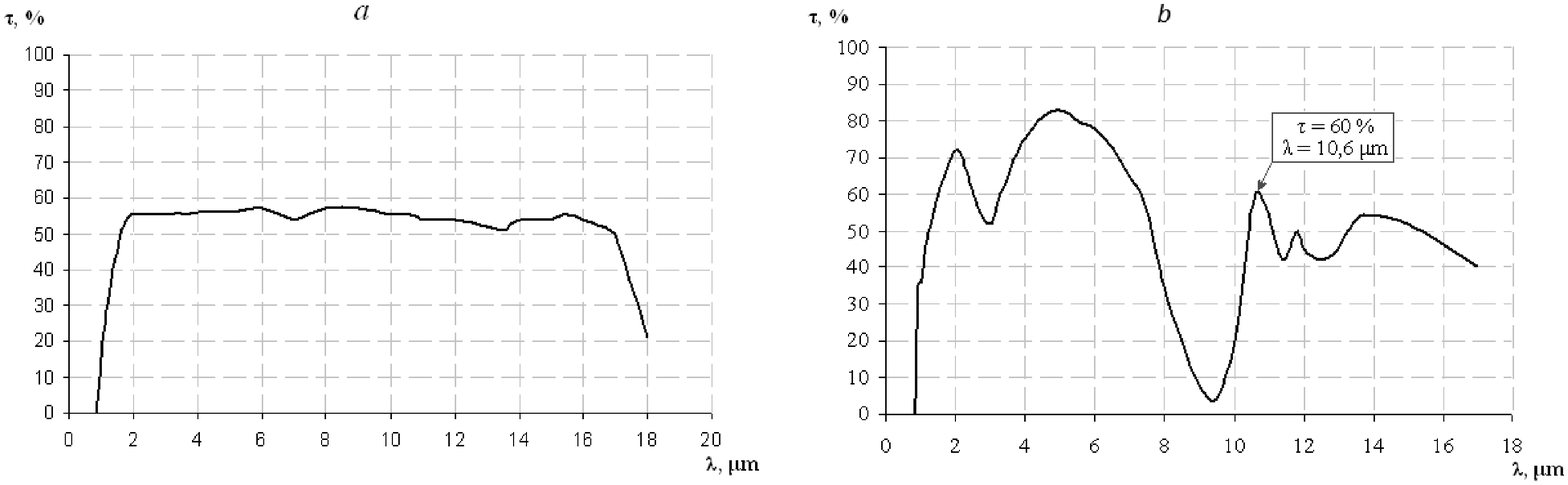}
\caption{The transmission spectra of $GaAs$ are shown for a) the
         3 mm thick original plate; b) the
         plate covered by SiO2 film with thickness of 0.6 $\mu$m.}
\label{asgaopic}
\includegraphics[scale=0.35]{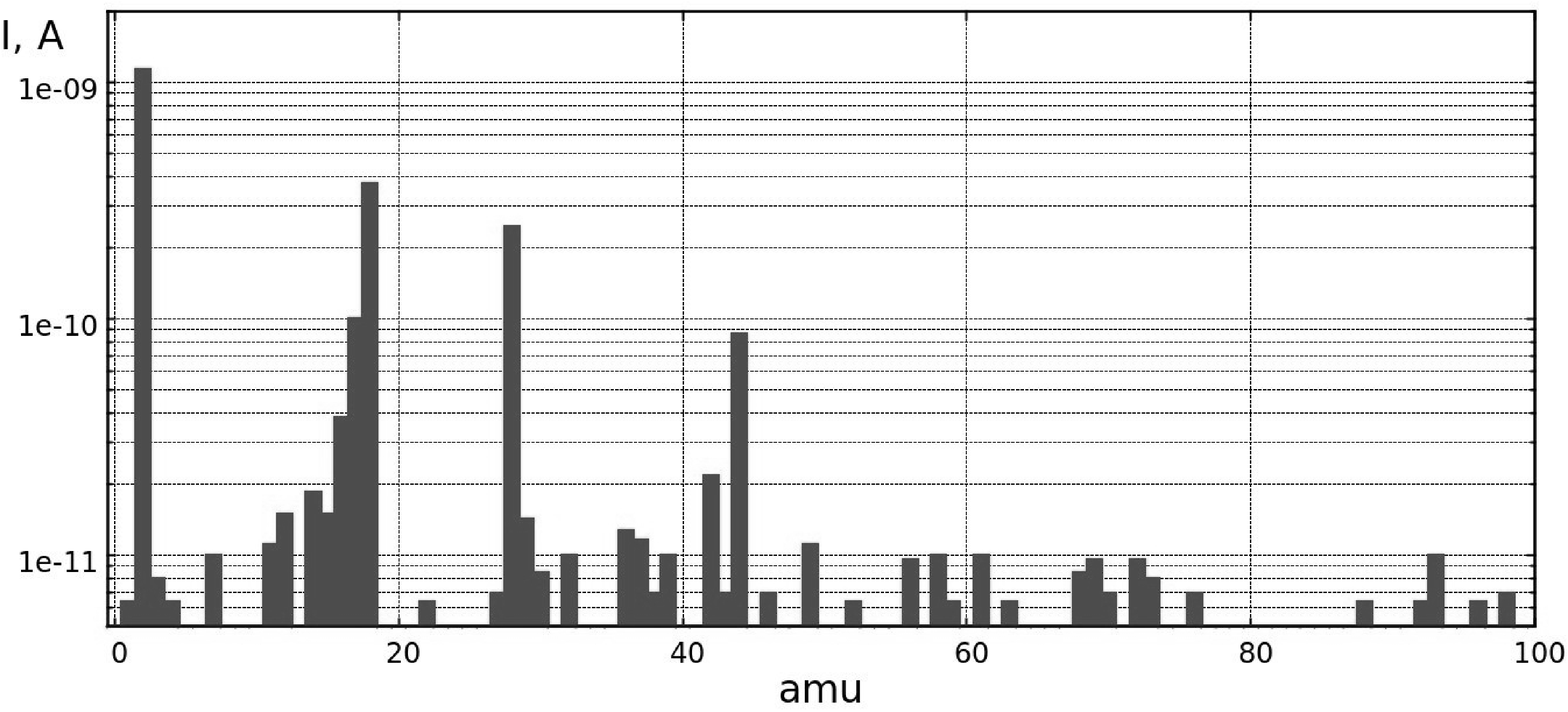}
\caption{Residual gas spectrum.}
\label{vacuum}
\end{center}
\end{figure}

 The viewport based on the GaAs mono-crystal provides: 
\begin{enumerate}
\item
 transmission spectrum from 0.9 up to 18 $\mu$m,
\item
 baking out of the vacuum system up to 250$^\circ$C,
\item
 extra high vacuum.
\end{enumerate}
 The viewport design is shown in Fig.~{\ref{viewport}}. It includes a
 304 L steel DN63 conflat flange and a GaAs crystal plate with
 diameter of 50.8 mm and thickness of 3 mm. In order to compensate
 mechanically for the difference of the GaAs and stainless steel
 thermal expansion coefficients, the GaAs plate is brazed with pure
 soft lead to a titanium ring, which in turn is brazed with AgCu alloy
 to the stainless steel ring. The stainless steel ring is welded to
 the flange. To avoid decomposition of the GaAs plate during brazing,
 it is covered with a 0.6 $\mu$m thick $SiO_2$ film using gas-phase
 deposition.  The transmission spectra of the plate before and
 after covering are shown in Fig.~\ref{asgaopic}. The transmission of
 the plate increases from 55 to 60 \% at the $CO_2$ laser wavelength
 $\lambda=10.6 \mu$m and from 20 to 35\% at $\lambda=1 \mu$m.

 After installation at BEPC-II, the vacuum chambers were baked out at
 $250^\circ$C for 24 hours. A pressure of $2\times 10^{-10}$ Torr was
 obtained. The residual gas spectrum is shown in Fig.~\ref{vacuum}.  

\subsection{Adjustment of the optical elements.}

The optical elements of the system were adjusted using SR. The copper
mirrors of the vacuum chambers and the mirrors of the optical system
were adjusted in such a way, that the SR light comes to the laser
output window. Actually the GaAs is not transparent for visible light
(Fig.~\ref{asgaopic}) but transmits infrared radiation. In
order to detect the infrared light, IR-sensitive video cameras were
used.

\subsection{HPGe detector}
 
 The purpose of a HPGe detector is to convert gamma rays into electrical
 impulses which can be used with suitable signal  processing, to determine
 their energy and intensity. A HPGe detector is a large germanium diode of the
 p-i-n type operated in the reverse bias mode. At a suitable operating 
 temperature (normally $\simeq$100 K), the barrier created at the junction 
 reduces the leakage current to acceptably low values. Thus an electric field 
 can be applied that is sufficient to collect the charge carriers liberated by
 the ionizing radiation.
 
 For the BEPC-II energy calibration system, we use the coaxial HPGe detector
 manufactured by ORTEC (model GEM25P4-70). It has diameter of 57.8
 mm and  height of 52.7 mm with 31.2\% relative efficiency\footnote{The
 efficiency of each detector is usually specified by a parameter called
 {\em relative detection efficiency}. The {\em relative detection efficiency}
 of coaxial germanium detectors is defined at 1.33 MeV relative to that of a 
 standard 3-in.-diameter, 3-in.-long $NaI(Tl)$ scintillator.}. The energy 
 resolution for the 1.33 MeV line of $^{60}Co$ is 1.74 keV (FWHM). The 
 detector is connected to the multi-channel analyzer ORTEC DSpec Pro (MCA), 
 which  transfers data using the USB port of the  computer.
 
 The HPGe spectrum has $2^{14}=16384$ channels. The bin error
 for each channel is defined as
\begin{equation}
 \Delta N =\sqrt{N+(\zeta N)^2},
\end{equation}
 where $N$ is number of counts in the channel and $\zeta$ corresponds
 to the MCA differential non-linearity, which is $\zeta=0.02$ according
 to the MCA specifications.

 Since the HPGe detector is located near the collider's beam pipes,
 background due to beam loss is extremely high.  In order to protect
 the HPGe detector from background, it is surrounded by 5 cm of lead
 on the sides, by 1.5 cm of iron below, and by 5 cm of lead above. The
 detector is also shielded by 10 cm of paraffin on all sides. Since
 the main background comes from the beam direction, an additional 11
 cm of lead is installed in these directions. Another 10 cm of lead
 can be moved into the beam using movable stages to shield from the
 beam direction that is not being measured and moved out when the
 beam is being measured.

\section{Data Acquisition System}

\begin{figure}[h]
\centering
\includegraphics*[width=1.\textwidth]{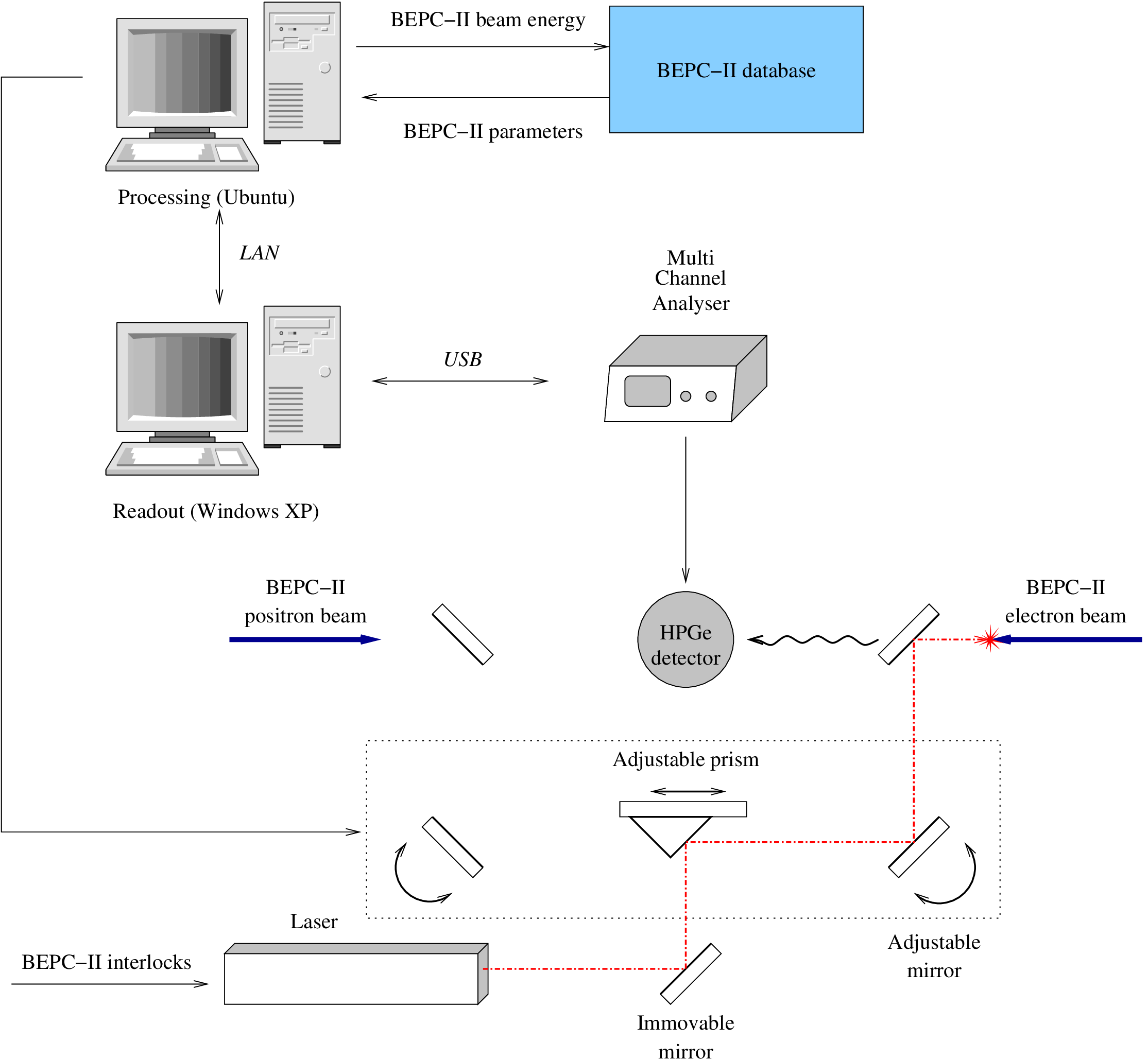}
\caption{Layout of data acquisition system.}
\label{das_layout}
\end{figure}

 The BEMS data acquisition system is shown in
 Fig.~\ref{das_layout}. The MCA digitizes the signal from the HPGe
 detector and produces the energy spectrum. It is connected to a
 Windows PC. All spectra processing, monitoring, and control over the
 devices involved in the BEMS is concentrated in another PC, under the
 control of Linux.
      
The data acquisition procedure is as follows. The HPGe detector
measurements are read every few seconds, and the detector counting
rate is calculated. If the requested acquisition time has elapsed, or
if conditions of the spectrum acquizition changed sufficiently, 
the current spectrum is saved to
a file and the next spectrum acquisition cycle is launched.
Simultaneously, another process periodically requests information from
the BEPC-II database and writes the BEPC-II parameters, such as beam
currents, lifetimes, and luminosity, to the file.

After finishing the spectrum acquisition cycle, another program
processes the spectrum; it calibrates the energy scale, finds the
Compton edge, and calculates the beam energy. The beam energy is
written into the BEPC-II database.  Since the BEPC-II parameters and
the detector counting rate are saved during the spectra acquisition,
conditions of any acquired spectrum can be analyzed at any time.

During data taking, mirrors are adjusted automatically to provide
maximal photon/electron (positron) interaction efficiency, using the
feedback from the detector counting rate.  The prism directing the
laser beam to either the electron or positron beams is controlled by
the same program, as are the movable stages that move the extra lead
protection in and out of the beam. The processing of the beam energy
measurement is fully automated by a script controlling the mirrors,
the prism, and the movable shielding.

\section{Data processing}

\begin{figure}[htbp]
\begin{center}
\includegraphics*[width=0.85\textwidth]{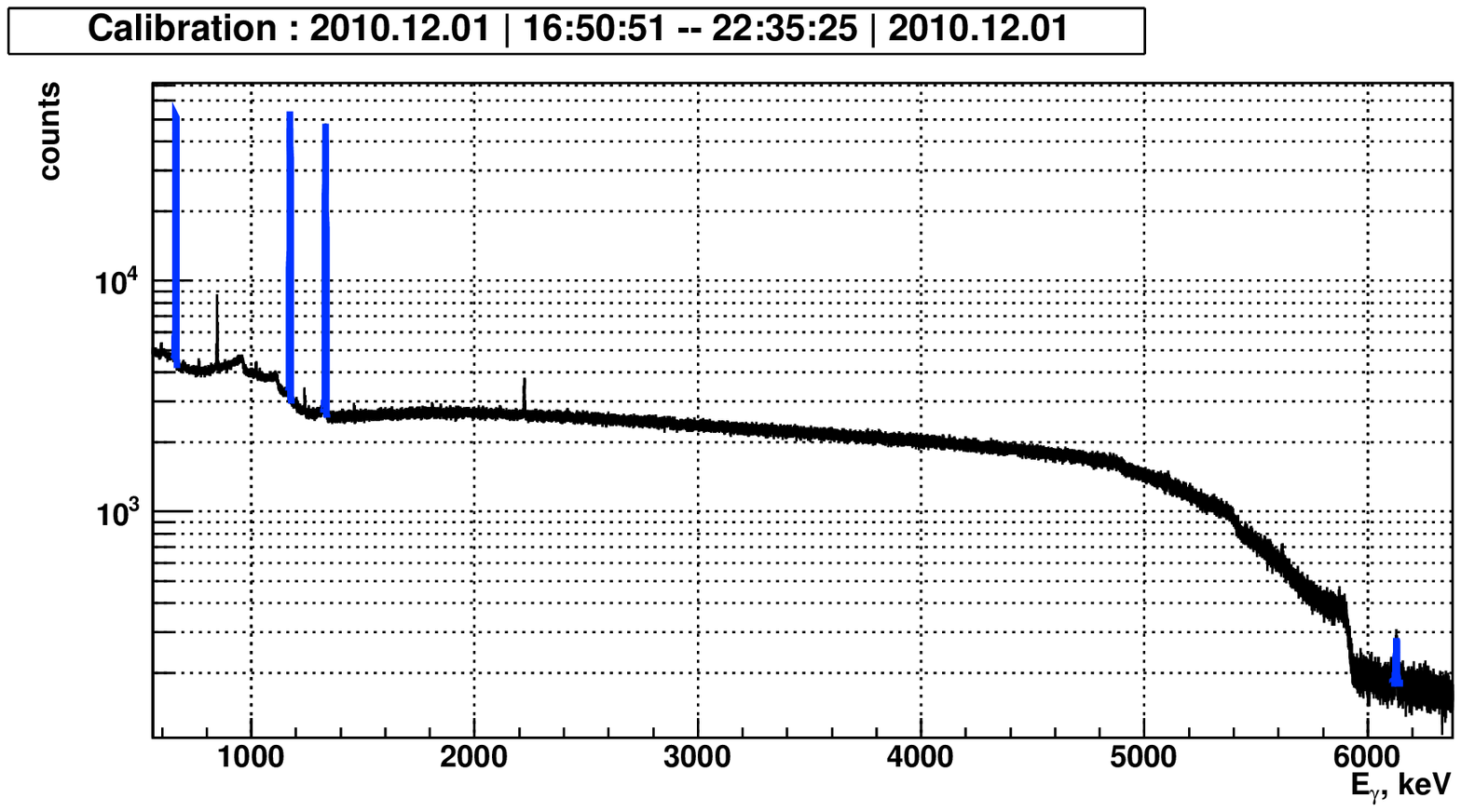}
\caption{The energy spectrum detected by the HPGe detector is shown.
 Several peaks, corresponding to  monochromatic $\gamma$-radiation  
 radiative sources, as well as the abrupt edge of the Compton photons spectrum 
 slightly below 6000~keV are clearly seen.}
\label{spectrum}
\end{center}
\end{figure}

 The processing of the spectrum (Fig.~\ref{spectrum}) includes
 calibration of the energy scale, Compton edge fitting and the
 calculation of the beam energy.

 The energy scale must be calibrated in the range from 2 to 8
 MeV. This is the energy range of back-scattered photons at BEPCII
 with beam energies from 1000 to 2100 MeV (Fig.~\ref{lines}).  The
 following sources were used in this work:
\begin{itemize}
\item 
 $^{137}$Cs      :  $E_\gamma = 661.657  \pm 0.003$~keV
\item 
 $^{60}$Co       :  $E_\gamma = 1173.228 \pm 0.003$~keV
\item 
 $^{60}$Co       :  $E_\gamma = 1332.492 \pm 0.004$~keV
\item 
$^{16}$O$^*$    :  $E_\gamma = 6129.266 \pm 0.054$~keV
\footnote{The [$^{238}$Pu~$^{13}$C] gamma source is used.
 The nuclear reaction occurs in this source:
 $\alpha + ^{13}\!\mbox{C} \rightarrow n + ^{16}\!\mbox{O}^*$.
 The excited oxygen emits $\gamma$-rays with energy of
 6129.266 $\pm$ 0.054~keV.~\cite{Alkemade1982383}}
\end{itemize}
 
 The goal of HPGe detector calibration is to obtain the coefficients
 needed for conversion of the HPGe detector's ADC counts into
 corresponding energy deposition, measured in units of keV, as well as
 the determination of the detector's response function parameters. The
 following response function is used:

\begin{equation} 
 f(x,x_0,\sigma,\xi) = {N\over\sqrt{2\pi}\sigma} \cdot
\left\{\begin{array}{ll}
  \exp\biggl\{ {-{(x-x_0)^2\over 2\sigma^2}} \biggr\}, 
  & x > x_0-\xi\cdot\sigma \\
  \exp\biggl\{ {\xi^2\over 2}{+{\xi(x-x_0)\over \sigma}} \biggr\}, 
  & x \leq x_0-\xi\cdot\sigma,  \\
\end{array}\right. 
\label{m-gauss-norm}
\end{equation}
\begin{equation}
 {1\over N} = \int_{-\infty}^{+\infty} f(x,x_0,\sigma,\xi) dx = 
 {1\over2}\mathrm{erfc}\biggl(-{\xi\over\sqrt{2}}\biggr)+
 {1\over\sqrt{2\pi}\xi}\exp\biggl(-{\xi^2\over 2}\biggr).
\end{equation}
 Here $x_0$ is the position of the maximum, $\xi$ is an asymmetry
 parameter, and $\sigma$ is the full-width of the Gaussian distribution at
 half-maximum divided by 2.36.

% The high stability of apparature -- HPGe and MCA, provides to perform the
% linear pre calibration of the HPGe energy scale using $^{137}Cs$ and 
% $^{60}Co$ rediative sources before long term operation of the BEMS. Thus
% the energy spectra scale is in the units of keV from the very begining and
% that is convenient for the further data processing.
 
The calibration procedure is as follows:
\begin{enumerate}
\item
 Peak searching is performed using a ROOT algorithm based on
 Refs.~\cite{nouck-1,nouck-2,nouck-3};
\item
 The found peaks are identified using the atlas of the well known radiative 
 lines;
\item
 The peaks which correspond to calibration lines are fitted
 by the sum of signal and background
 distributions,  $f(x,x_0,\sigma,\xi)+p_1(x)$ (Fig.~\ref{O16}), where $p_1(x)$ 
 is a first-order polynomial. The free parameters of the fit are $x_0$, 
 $\sigma$,  $\xi$, and the coefficients of the polynomial.
\begin{figure}[htbp]
\centering
\includegraphics[width=0.85\textwidth]{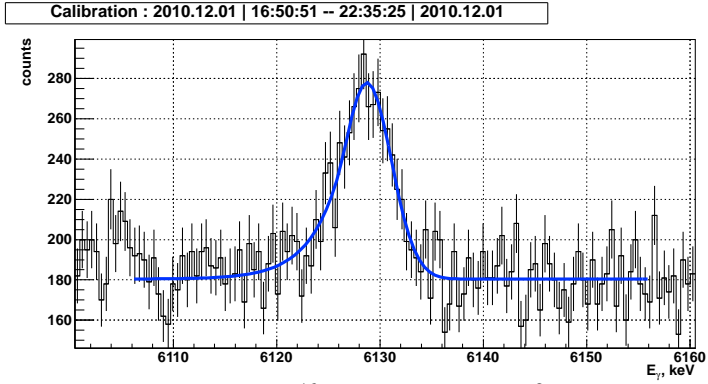}
\vspace{-5mm}
\caption{The fit to the $^{16}O^*$ 6.13~MeV peak. $\chi^2/NDF$ = 87.2/105}
\label{O16}
\end{figure}
\item
 Using  the fit results, the energy dependencies of the response
 function (Eqn.~\ref{m-gauss-norm}) parameters are determined.
 The $\sigma$ energy dependence (Fig.~\ref{eres}) is described by the formula:
\begin{equation}
 \sigma_E = \sqrt{ K_0 + F  E_\gamma}\;,
 \label{real_resolution}
\end{equation}
 where $E_\gamma$ is the photon energy, $K_0$ = 0.772 $\pm$ 0.020
 keV$^2$, and $F$ = 0.56 $\pm$ 0.02 eV. The energy dependence of the
 asymmetry parameter $\xi$ is approximated with an empirical
 function, $g(x) = p_0 + p_1 \exp (-p_2 x)$ (Fig.~\ref{tail}).  In
 order to obtain the correction to the measured energy due to spectrometer
 scale non-linearity, the difference between positions of the
 calibration peaks $x_0$ and their known reference values are fitted by
 a second-order polynomial (Fig.~\ref{nonl}).
\begin{figure}[htbp]
\begin{center}
\includegraphics[width=0.8\textwidth]{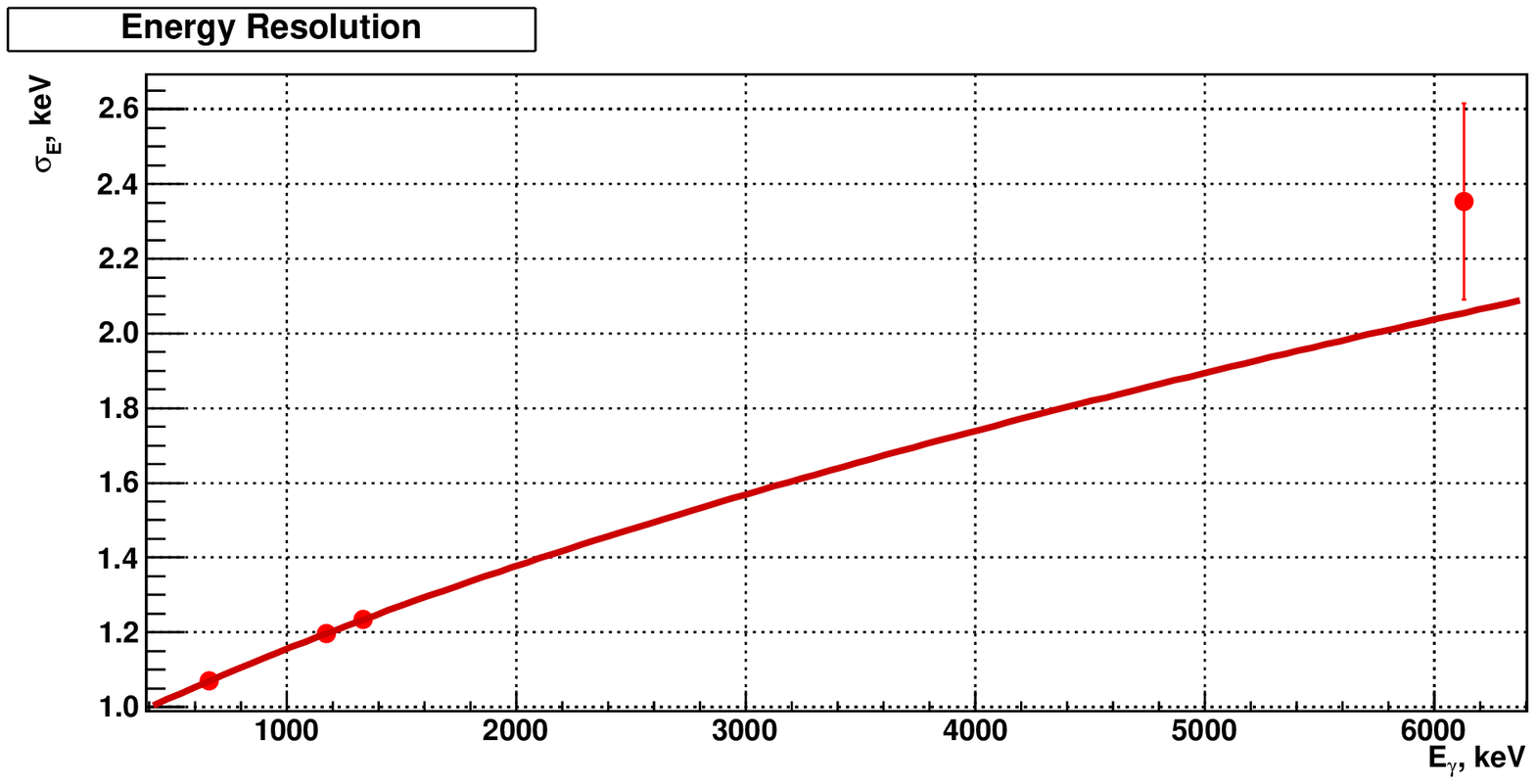}
\caption{$\sigma_E$ vs the photon energy, fitted by Eqn.~
\ref{real_resolution}. The fit results in $\chi^2/NDF$ = 1.3/2.}
\label{eres}
\includegraphics[width=0.8\textwidth]{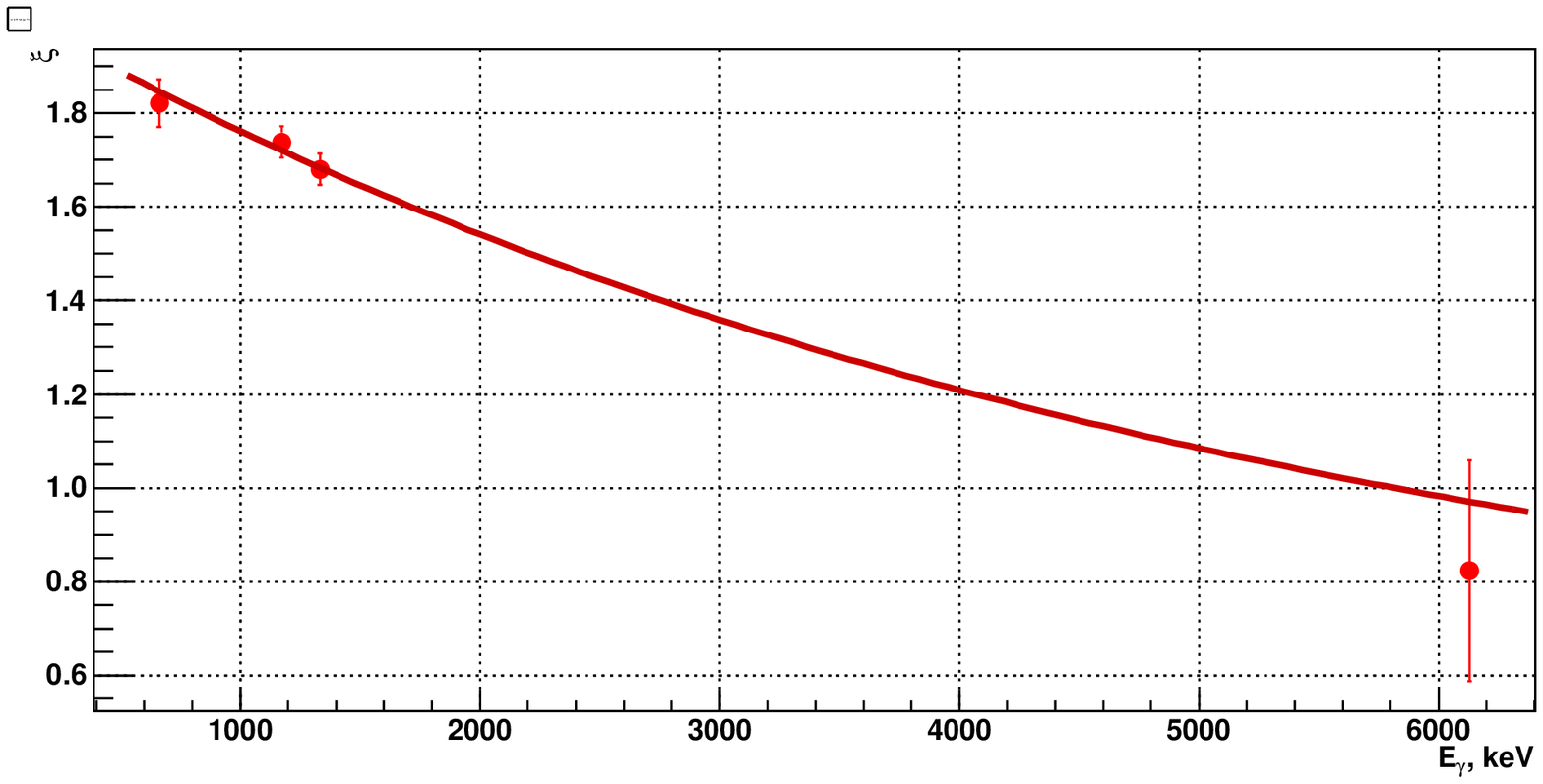}
\caption{Asymmetry parameter $\xi$ vs photon energy. }
\label{tail}
\includegraphics[width=0.8\textwidth]{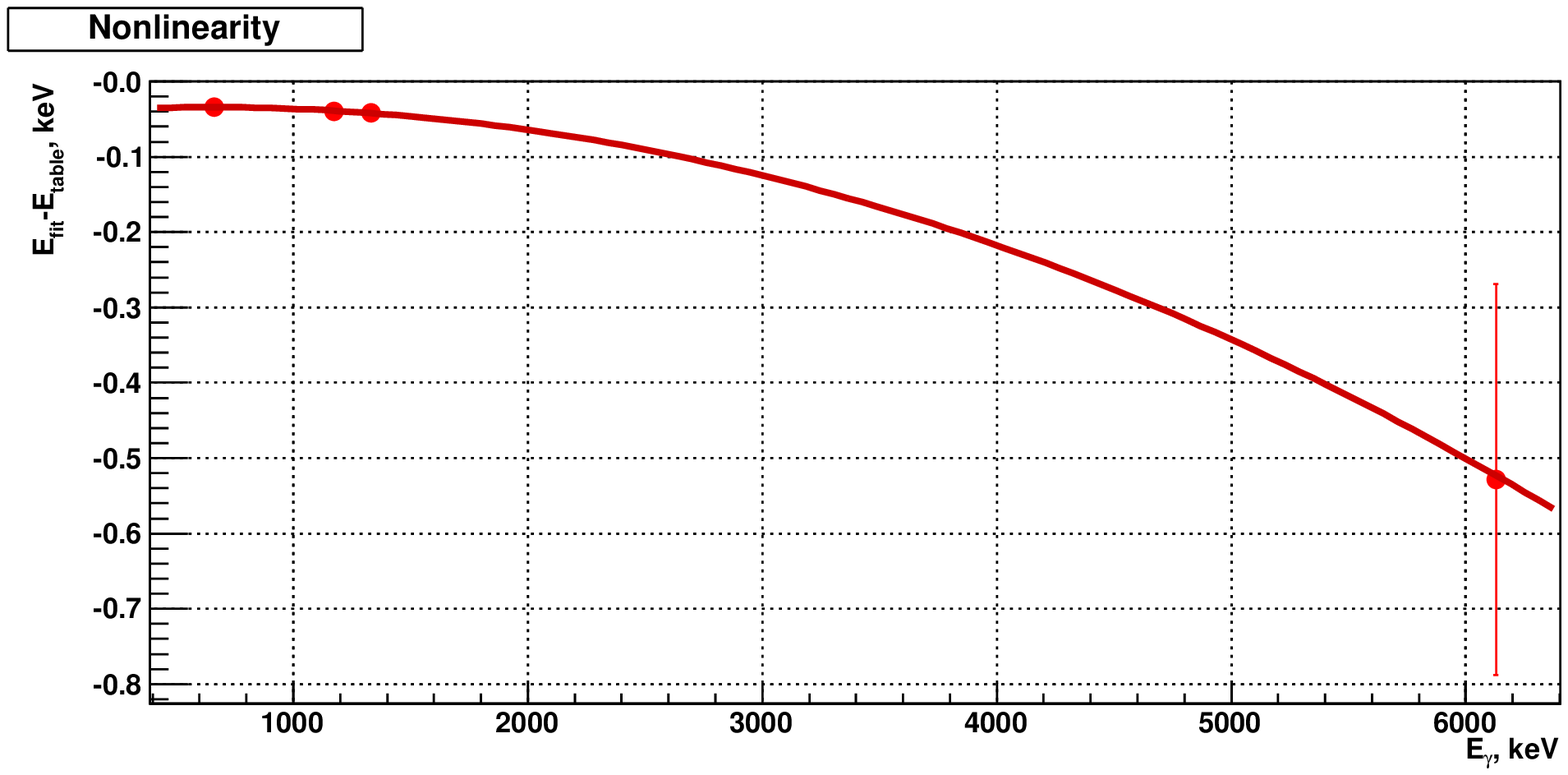}
\caption{Energy dependence of the differences between the calibration peaks 
%energy obtained using linear pre calibration,
from their true values.}
\label{nonl}
\end{center}
\end{figure}
\end{enumerate}

\begin{figure}[htbp]
\begin{center}
\includegraphics[width=0.8\textwidth]{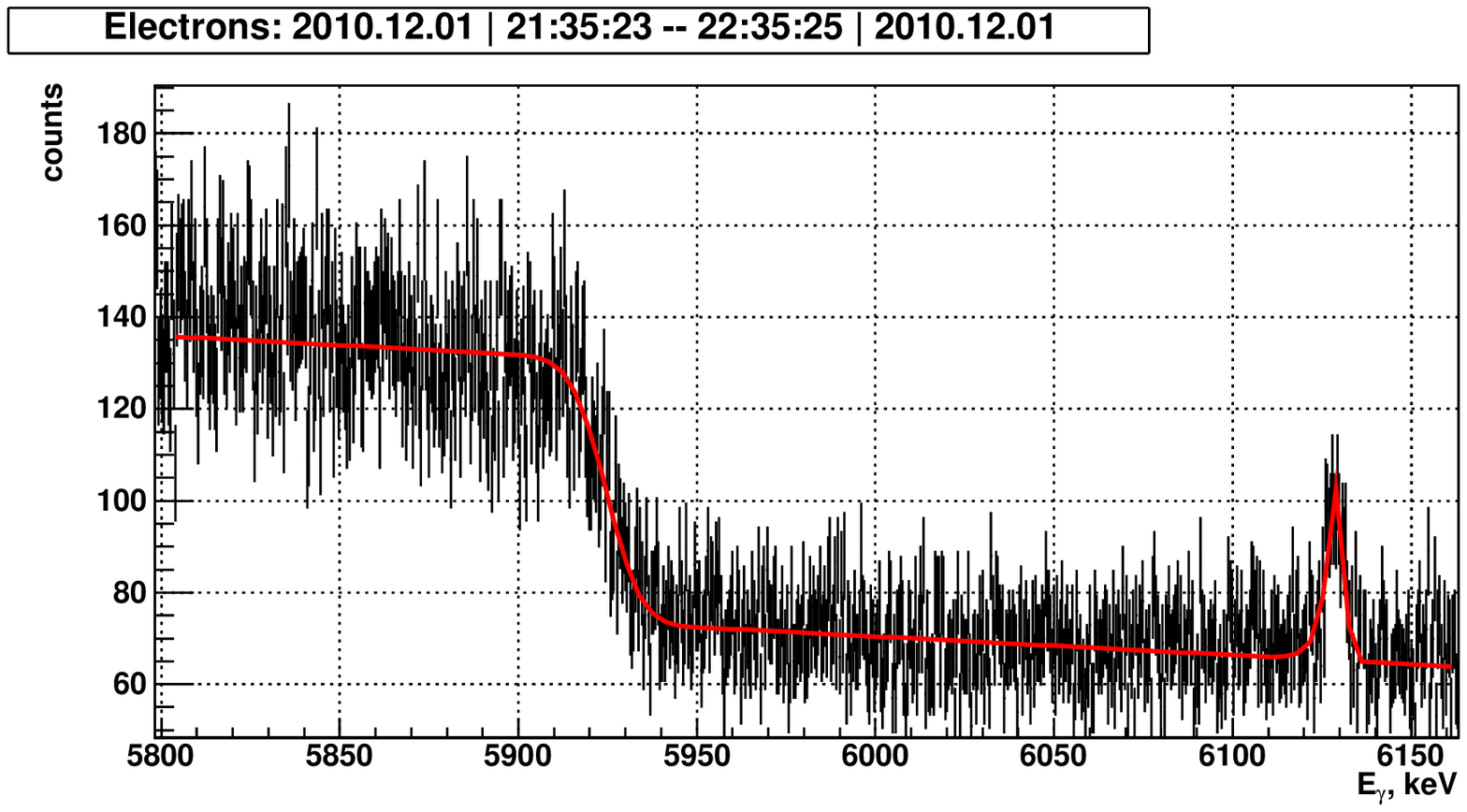}
\caption{The energy spectrum of back-scattered photons near $\omega_{max}$ and
the fit function. The 6.129 MeV peak is also seen.}
\label{edge-2}
\end{center}
\end{figure}
 The edge of the back-scattered photon spectrum (Fig.~\ref{edge-2}) is fitted  by 
 the function:
\begin{equation}
S_2(x,x_0,\sigma,\sigma_s,\xi) = \int\limits_{x}^{+\infty}
S_1(y,x_0,\sigma,\sigma_s,\xi)\;dy\; +p_1(x).
\label{convolution-5}
\end{equation}
 Here $p_1(x)$ takes into account the background contribution, and $S_1$ is
 a convolution of the step function $\theta(x_0-x)$:
\begin{equation}
\theta(x_0-x) = 
\left\{\begin{array}{ll}
  0,  & x < x_0  \\
  1, &  x > x_0,  \\
\end{array}\right. 
\end{equation}
 which describes the ``pure'' edge shape with the HPGe detector
 response function (Eqn.~\ref{m-gauss-norm}) and a Gaussian:
\begin{equation}
 g(x,x_0,\sigma_s) = {1\over \sqrt{2\pi}\sigma_s} 
 \exp\biggl\{ -{{(x-x_0)^2}\over{2\sigma_s^2}}\biggr\},
\end{equation}
 which takes into account the energy spread of  back-scattered photons due to
 energy distribution of the collider beam.
\begin{eqnarray}
\nonumber & S_1(x,x_0,\sigma,\sigma_s,\xi) =
\displaystyle\frac{N}{2\sqrt{2\pi}} \times &  \\
& \times \Biggl[  \displaystyle\frac{1}{\sigma}
\exp\Bigl(\frac{\xi^2}{2}\bigl(1+\frac{\sigma_s^2}{\sigma^2}\bigr)+
\frac{\xi x}{\sigma}\Bigr)
\cdot\mathrm{erfc}\Bigl(\frac{\xi(\sigma^2+\sigma_s^2)+x\sigma}
{\sqrt{2}\sigma\sigma_s}\Bigr) + &\\
\nonumber & +  \displaystyle\frac{1}{\sqrt{\sigma^2+\sigma_s^2}}
\exp\Bigl(-\frac{x^2}{2(\sigma^2+\sigma_s^2)}\Bigr)
\cdot\mathrm{erfc}\Bigl(-\frac{\xi(\sigma^2+\sigma_s^2)+x\sigma}
{\sqrt{2(\sigma^2+\sigma_s^2)}\sigma_s}\Bigr) \Biggr] & .
\label{convolution-4}
\end{eqnarray}
The edge position $\omega_{max}\equiv x_0$, $\sigma_s$, and the coefficients of the 
first-order polynomial $p_1(x)$ are the free parameters of the fit. 
Using the $\omega_{max}$ value obtained from the fit, the average beam energy
$\varepsilon_{nip}$ ($nip$ denotes north interaction region) in the $e-\gamma$ 
interaction region is calculated according to formula (\ref{nepec}).
Taking into account the energy losses due to synchrotron radiation, the beam
energy in the south interaction point ($sip$) is obtained as
\begin{equation}
 \varepsilon_{sip}(\mathrm{MeV})  = \varepsilon_{nip}(\mathrm{MeV}) +
 4.75\cdot10^{-3}*(0.001 \cdot \varepsilon_{nip}(\mathrm{MeV}))^4.
\label{cme}
\end{equation}

\section{System performance}

 The system was put in operation and tested with beams of energy about
 1840 MeV.  The relative statistical accuracy of the beam energy
 determination of about $5 \cdot10^{-5}$ was achieved after
 approximately 1 hour of data taking.  The systematical accuracy was
 studied by comparison of the well known mass of the $\psi^\prime$
 resonance $m_{\psi^\prime}=3686.09\pm 0.04$~MeV \cite{PDG} with its
 value obtained using the BEMS.
 
 In order to obtain the $\psi^\prime$ mass two scans of the resonance
 energy region were done with a total integrated luminosity of about
 $3.95~\mbox{pb}^{-1}$. The data were collected at 12 energy points
 over 36 hours.  The $\psi^\prime$ mass was measured as follows.
\begin{enumerate}
\item
 The multihadronic $e^+e^-\to hadrons$ events were selected.
\item
 The events of $e^+e^-\to\gamma\gamma$ were used to determine the integrated 
 luminosity $L$:
\begin{equation}
 L= {N^{\gamma\gamma} \over \sigma^{\gamma\gamma}(w)},
\end{equation}
 where $N^{\gamma\gamma}$ and $\sigma^{\gamma\gamma}$ are the selected
 number of events and cross section obtained using Monte Carlo
 simulation, and $w$ is the center of mass energy.
\item
 The resonance mass was obtained from the fit of the number of
 $e^+e^-\to hadrons$ events expected, $M^{mhad}=\sigma^{mhad} L$, to the
 number of detected multihadronic events $N^{mhad}$. Here
 $\sigma^{mhad}$ is the expected cross section of
 $e^+e^-\to hadrons$:
\begin{equation}
\sigma^{mhad}(w) = \sigma_{bg} \cdot \left(\frac{3686\,
\mbox{MeV}}{w}\right)^2 +
\epsilon \cdot \sigma_{res}(w,m,\sigma_w),
\end{equation}
 where $m$ is the $\psi^\prime$ meson mass, $\sigma_{bg}$ is the
 nonresonant background cross section, $\epsilon$ is the detection
 efficiency, $\sigma_{res}$ is the cross section of the $\psi^\prime$
 resonance production $\sigma_0(w,m)$ \cite{CS} convoluted with the
 beam energy spread $\sigma_w$:
\begin{equation}
 \sigma_{res}(w,m,\sigma_W) = \int_{-\infty}^{+\infty}
 \frac{\exp\left(-\frac{(w-w^\prime)^2}{2\sigma_w^2}\right) }
 {\sqrt{2\pi}\sigma_w} \sigma_0(w^\prime,m) \; dw^\prime
\end{equation}
\end{enumerate}

 Charged tracks were selected requiring their point of closest approach to
 the beam axis be within 1 cm of the beam line, and their angle with respect
 to the beam axis, $\theta$, to satisfy $|\cos\theta|<0.93$ \cite{bes_an}.
 Photon candidates must have at least 25 (50) MeV of energy in the barrel 
 (end cap) electromagnetic calorimeter (EMC) and  have $|\cos\theta|<0.82$
 ($0.86 < |\cos\theta| < 0.92$).

 The $e^+e^-\to\gamma\gamma$  events were selected using the
 following criteria
\begin{enumerate}
\item
 $N_q=0$ and $N_\gamma>1$,  where $N_q$
 is the number of charged tracks and $N_\gamma$ is the number of photons;
\item
 $|\cos \theta_i|<0.8$, where here and below  $i=1,2$ denotes the photons with
 the highest energy deposition;
\item $|\Delta\theta|=|\pi-(\theta_1+\theta_2)|<0.05$;
\item $-0.06 < \Delta \phi <0.02$, $\Delta\phi=\pi-|\phi_1-\phi_2|$,
  where $\phi$ is the azimuthal angle around the beam direction;
\item $0.8<E_i/E_{beam}<1.2$, where $E_i$ is the energy deposition in
 the EMC of
 the $i$th photon and $E_{beam}$ is the beam energy.
\end{enumerate}

 In order to select multihadronic events the following criteria were
 applied.
\begin{enumerate}
\item
 $N_q>3$;
\item
 $S>0.06$, where  $S = \frac{3}{2}(\lambda_2 + \lambda_3)$ is the sphericity
 parameter. Here $\lambda_1\ge\lambda_2\ge\lambda_3$ are
 eigenvalues of sphericity tensor:
 \[ S^{ij} = \frac{\sum\limits^{N_q}_{n=1} p^i_n p^j_n}
 { \sum\limits_{n=1}^{N_q} p_n^2 },  \]
 where $p_i$ the momentum of the $ith$ track.
\end{enumerate}
		       
 The number of selected multihadronic events $N^{mhad}$ were fitted by  
 minimizing the likelihood function:
\begin{equation}
 \chi^{2} =\sum\limits_{i=1}^{N}
 \frac{(N^{mhad}_i-\sigma^{mhad} L_i)^2}
 {N^{mhad}_i(1+N^{mhad}_i/N^{ee,\gamma\gamma}_i)} +
 \sum_{i=1}^{N} \left( \frac{w_i - W_i}{\Delta W_i} \right)^2.
\end{equation}
 The free parameters of the fit were $\psi^\prime$ mass $m$,
 $\sigma_{bg}$, $\epsilon$, $\sigma_w$. The center of mass energy $w_i$ at 
 each energy point was fitted to the values $W_i$ obtained as follows:
\begin{equation}
W = 2 \sqrt{\varepsilon_{sip}^- \varepsilon_{sip}^+} \cos\frac{\alpha}{2},
\end{equation}
where $\varepsilon_{sip}^-$ and $\varepsilon_{sip}^+$ are the energies
of the electron and positron beams respectively in the south
interaction region calculated according to formula (\ref{cme}),
$\alpha=22$~mrad is the crossing angle of the beams. The error $\Delta
W$ of the W determination is calculated from the errors of
$\varepsilon_{sip}^-$ and $\varepsilon_{sip}^+$.

\begin{figure}[htp]
\begin{center}
\includegraphics[width=0.7\textwidth]{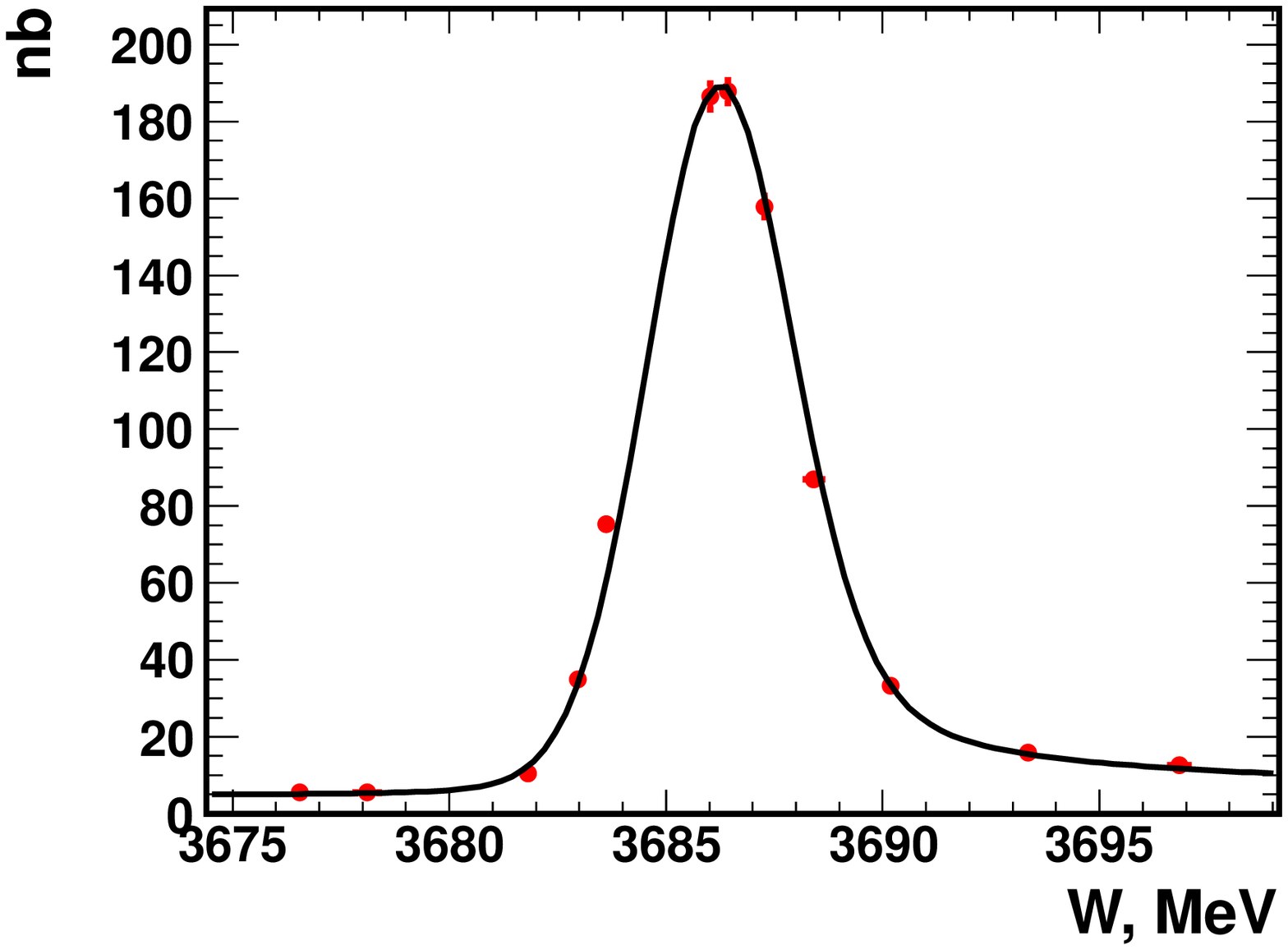}
\caption{Fit of the $\psi^\prime$.}
\label{ggfit}
\includegraphics[width=0.7\textwidth]{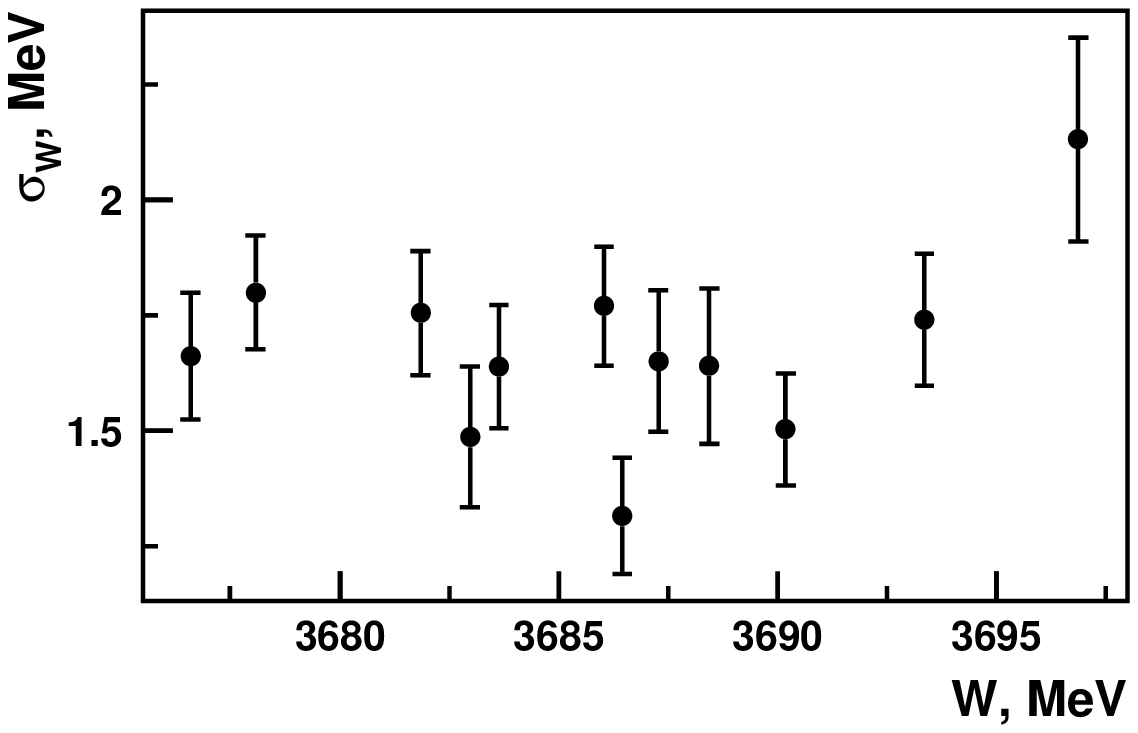}
\caption{The center-of-mass energy spreads obtained by means of the BEMS.}
\label{pa-oc}
\end{center}
\end{figure}

The results of the fits for the two scans are in agreement. The results of
the fit to all data are presented in Table~\ref{tab2} and in Fig.~\ref{ggfit}.
\begin{table}[h!]
\caption{The results of the fit. $\Delta m = m - m_{\psi^\prime}$.}
\begin{center}
\begin{tabular}{lc} \hline
$\Delta m$ (keV)      & $1\pm56$ \\
$\sigma_w$ (MeV)      & $1.58\pm0.03$   \\
$\sigma_{bg}$ (nb)    & $4.7\pm 0.1$    \\
$\varepsilon$ (\%)    & $32.7\pm0.5$    \\
$\chi^2/ndf$         & $13.5/8$  \\
$P(\chi^2,ndf) (\%) $ & $9.7$ \\ \hline
\end{tabular}
\end{center}
\label{tab2}
\end{table}

 In order to check the adequacy of the selection criteria for
 multihadron events, even more strict cuts were applied for their
 selection:
\begin{itemize}
\item $N_q>4$
\item $p_t>50$~MeV and $|\cos\theta|<0.8$ for each charged track. Here $p_t$
 is the transverse momentum.
\end{itemize}
 The fit was performed with the new number of selected multihadronic
 events, and $\Delta m=-17 \pm 58$ keV and $\sigma_w=1.56\pm 0.03$ MeV
 were obtained.

 The luminosity determination was also tested using events of
 $e^+e^-\to e^+e^-$, scattered at small angles to suppress the
 contribution from $\psi^\prime\to e^+e^-$ decay. The mass difference
 and beam energy spread were found to be $\Delta m=17 \pm 50$ keV and
 $\sigma_w=1.59\pm 0.03$ MeV.

The bias of the center of mass energy obtained using the BEMS from the
true value can be estimated as $\Delta m =m - m_{\psi^\prime}$:
\begin{equation}
\Delta m = 1\pm 56\pm 24\pm 40 \simeq 1\pm 72 \mbox{~~keV}.
\end{equation}
Here the first error is the statistical, the second is due to systematical
uncertainties of the luminosity determination and multihadronic event
selection, and the last one is the error of the PDG $m_{\psi^\prime}$ value.
If we take the deviation of the measured beam energy $\Delta E$ from
the actual value as $\Delta\varepsilon = \Delta m / 2$, then
\begin{equation}
\Delta\varepsilon = 1\pm 36 \mbox{~~keV}
\end{equation}
 Taking into account this deviation, the relative accuracy of the beam energy 
 determination can be estimated as
 $2\cdot 10^{-5}$.

 The center-of-mass energy spreads $\sigma_W$ obtained by using the
 BEMS at the 12 energies are shown in Fig.~\ref{pa-oc}. The relative
 statistical accuracy of the $\sigma_W$ determination per measurement
 is about 10\%. The average of all measurements is $\sigma_W=1.65\pm
 0.04$. The energy spread obtained from the fit of $\psi^\prime$
 resonance is $\sigma_w=1.58\pm 0.03$. The difference between these
 two values $\sigma_W-\sigma_w=0.07\pm0.05$ is about 1.4 standard
 deviations and consistent with zero. Using this difference, the
 relative systematical accuracy of the energy spread determination can
 be estimated as 6\%.
 
\section{Conclusion}
 
  The energy measurement system of the BEPC-II collider beams based on
  the Compton back-scattering method was designed, constructed, and put
  into operation.  The systematical error of the beam energy
  determination is tested through measurement of the $\psi^\prime$
  mass and is estimated as $2\cdot 10^{-5}$.

\section*{Acknowledgment}
  The authors are grateful to A.E. Bondar, E.B. Levichev, Yu.A. Tikhonov for
  initiating and supporting the work. The work was supported in part by
  SB RAS joint project No. 32 for fundamental research with CAS;
  National Natural Science Foundation of China (10775412, 10825524,
  10935008), Instrument Developing Project of Chinese Academy of
  Sciences (YZ200713), Major State Basic Research Development Program
  (2009CB825200, 2009CB825203, 2009CB825206) and Knowledge Innovation
  Project of Chinese Academy of Sciences (KJCX2-YW-N29); and by the
  Department of Energy under Contract No. DE-FG02-04ER41291 (u. of Hwasii).


\begin{thebibliography}{9}
\bibitem{bepc2}
J.Q. Wang, et al., Proceedings of IPAC'10, Kyoto, Japan, (2010) 2359
\bibitem{bes3}
M. Ablikim, et al., Nucl. Instr. Meth. A 614 (2010) 345
\bibitem{yellbook}
D. M. Asner, et al., arXiv:0809.1869
\bibitem{PDG}
K. Nakamura et al. (Particle Data Group), J. Phys. G 37, 
 075021 (2010)   
\bibitem{mo1}
Y.K. Wang, X.H. Mo, C.Z.Yuan, J.P. Liu, Nucl. Instr. and Meth
 A 583 (2007) 479.    
\bibitem{mo2}
X.H. Mo, In Proceedings of 9-th Intirnational Workshop on tau
 lepton physics, Pisa, Italy, September 19-22, 2006, Nucl. Phys. Proc. Suppl.
 169 (2007) 132    
\bibitem{rd} A.N. Skrinsky and Yu.M. Shatunov, Sov. Phys. Uspekhi 32 (1989) 548
\bibitem{bes-tau} J.Z. Bai et al., Phys. Rev. D 53 (1996) 20
\bibitem{okp1}
 T. Yamazaki et. al., IEEE Trans. on Nucl. Sci., Vol. NS-32, No5, 1985, p.3406
\bibitem{okp2}
 Ian C. Hsu et. al., Nucl. Instr and Meth. A 384 (1997) 307-315; \\
 Phys. Rev. E 54, 1996, 5657
\bibitem{okp3}
 R. Klein et al., Nucl. Instr. Meth. A 384 (1997) 293; \\
 J. Synchrotron Rad. 5 (1998) 392; \\
 Nucl. Instr. Meth. A 486 (2002) 545
\bibitem{okp4}
 N. Muchnoi et al., In Proc. of the EPAC, Scotland, Eidenburgh,June 26-30,
 2006, EPAC 1181; \\
 V.E. Blinov et al, in Proc. of International Conference on instrumentation
 for colliding beam physics,  Novosibirsk, Russia February 28 -
 March 5, 2008, Nucl. Instr. and Meth. A 598 (2009) 23 \\
 V.E. Blinov, et al., ICFA Beam Dyn. Newslett. 48 (2009) 195 \\
 O.V. Anchugov et al., Zh. Eksp.Teor. Fiz. 136 (2009) [J.Exp.Theor.Phys. 109
 (2009) 590]
\bibitem{proposal}
 M.N. Achasov et al., BINP Preprint 2008-4 (2008); ArXiv:0804.0159 (2008) \\
\bibitem{prop}
 M.N.Achasov, et al., in Proc. of the 10th Int. Workshop on Tau Lepton
 Physics, Novosibirsk, Russia, September 22-25, 2008,
 Nucl. Phys. Proc. Suppl. 189 (2009) 366 \\
 MO Xiao-Hu, et al., Proc. of the Int. Workshop on $e^+e^-$ collisions from 
 $\phi$ to $\psi$, October 13 - 16, 2009, Beijing, China, Chin. Phys. C 
 34(6) (2010) 912
\bibitem{CO2lambda}
 C.K.N. Patel, Phys. Rev 136(5A) (1964) 1187
\bibitem{win1}
 E.V.Abakumova et al., Vacuum Technic and Technology, 20(2) (2010) 77 
 (in Russian)
\bibitem{Alkemade1982383}
 P.F.A. Alkemade, C. Alderliesten, P. De Wit, and C. Van der Leun, Nucl.
 Instr. and Meth. A 197(2-3) (1982) 383-390 
\bibitem{nouck-1}
 Tibor Papp, X-Ray Spectrometry 32(6) (2003) 1097-4539
\bibitem{nouck-2}
 M.C. Lee, K. verghese, R.P. Gardner, Nucl. Instr and Meth. A 262 (1987)
 430-438
\bibitem{nouck-3}
 H. Siegert, H. Janssen, Nucl. Instr and Meth. A 286 (1990) 415-420
\bibitem{CS}
 K. Yu. Todyshev, The application Breit-Wigner form with radiative corrections
 to the resonance fitting  http://arxiv.org/pdf/0902.4100v3
\bibitem{bes_an}
 M. Ablikim, et. al., Phys. Rev. D 81, 052005 (2010)
\end{thebibliography}
\end{document}